\documentclass[aoas,preprint]{imsart}
\usepackage{lineno}

\usepackage{chngcntr}

\usepackage{algorithm,algorithmic,amsmath,amssymb,amsthm,etoolbox,graphicx,mathrsfs}
\usepackage[letterpaper, margin=1in]{geometry}
\usepackage[colorlinks]{hyperref}
\usepackage[symbol]{footmisc}
\usepackage{fixltx2e}
\usepackage{multirow}
\usepackage{chngcntr}
\RequirePackage{natbib}
\usepackage{enumitem}
\makeatletter
\def\namedlabel#1#2{\begingroup
    #2%
    \def\@currentlabel{#2}%
    \phantomsection\label{#1}\endgroup
}
\makeatother
\newcommand{\bel}{\begin{eqnarray}\label}
\newcommand{\eel}{\end{eqnarray}}
\newcommand{\bes}{\begin{eqnarray*}}
	\newcommand{\ees}{\end{eqnarray*}}
\newcommand{\bei}{\begin{itemize}}
	\newcommand{\eei}{\end{itemize}}

\newcommand\cites[1]{\citeauthor{#1}'s\ (\citeyear{#1})}

\newtheorem{lemma}{Lemma}
\newtheorem{prop}{Proposition}

\makeatother
\newcounter{parentnumber}

\theoremstyle{definition}

\def\<{\langle}
\def\>{\rangle}
\def\1{\mathbf{1}}

\def\argmax{\operatorname{argmax\ } \displaylimits}

\def\bb{\boldsymbol{\beta}}

\def\Cov{\mathrm{Cov}}

\def\E{\mathbb{E}}
\def\e{\mathbf{e}}
\def\g{\mathbf{g}}
\def\G{\Gamma}

\def\R{\mathbb{R}}
\def\S{\Sigma}

\def\s{\sigma}
\def\T{\top}

\def\tr{\mathrm{tr}}
\def\Var{\mathrm{Var}}
\def\v{\mathbf{v}}

\def\u{\mathbf{u}}
\def\y{\mathbf{y}}

\def\z{\mathbf{z}}

\makeatletter

\begin{document}
	\begin{frontmatter}
		
		\title{The Mahalanobis kernel for heritability estimation in genome-wide association studies: \\Fixed-effects and random-effects methods\protect\thanksref{T1}}
		\runtitle{The Mahalanobis kernel for heritability estimation}
		\thankstext{T1}{Supported by NSF Grant DMS-1454817.}
		
		\begin{aug}
			\author{\fnms{Ruijun}  \snm{Ma}\corref{}\ead[label=e1]{rma@stat.rutgers.edu}}
			\and \author{\fnms{Lee H.} \snm{Dicker}\ead[label=e2]{ldicker@stat.rutgers.edu}}

			\runauthor{R. Ma and L. H. Dicker}
			
			\affiliation{Department of Statistics and Biostatistics, Rutgers University, USA}
			
			\address{501 Hill Center, Piscataway, NJ, 08854\\ 
				\printead{e1,e2}}
		\end{aug}
		
		\begin{abstract}
			Linear mixed models (LMMs) are widely used for heritability estimation in genome-wide association studies (GWAS). In standard approaches to heritability estimation with LMMs, a genetic relationship matrix (GRM) must be specified.  In GWAS, the GRM is frequently a correlation matrix estimated from the study population's genotypes, which corresponds to a normalized Euclidean distance kernel. In this paper, we show that reliance on the Euclidean distance kernel contributes to several unresolved modeling inconsistencies in heritability estimation for GWAS.  These inconsistencies can cause biased heritability estimates in the presence of linkage disequilibrium (LD), depending on the distribution of causal variants. We show that these biases can be resolved (at least at the modeling level) if one adopts a Mahalanobis distance-based GRM for LMM analysis.  Additionally, we propose a new definition of partitioned heritability -- the heritability attributable to a subset of genes or single nucleotide polymorphisms (SNPs) -- using the Mahalanobis GRM, and show that it inherits many of the nice consistency properties identified in our original analysis.  Partitioned heritability is a relatively new area for GWAS analysis, where inconsistency issues related to LD have previously been known to be especially pernicious.  
		\end{abstract}
		
		
		
	\end{frontmatter}
	
\allowdisplaybreaks

\section{Introduction}

Heritability is the proportion of phenotypic variance explained by genetic variance \citep{falconer1960introduction,lynch1998genetics}. There are many different definitions of heritability and different methods for estimating heritability from data \citep[e.g.][]{haseman1972investigation,henderson1984applications,visscher2008heritability,yang2011gcta,golan2014measuring,bulik2015ld}.  This paper is focused on heritability estimation methods that are related to variance components estimation methods for linear mixed models (LMMs).  LMM-based methods for heritability estimation have been used since the 1950s \citep{henderson1950estimation}; additionally, over the last decade they have emerged as one of the most widely-used methods for estimating heritability with genome-wide association study (GWAS) data \citep{hindorff2009potential,yang2010common,kang2010variance,zaitlen2012heritability}.  However, standard approaches to heritability estimation with LMMs have some unresolved inconsistencies that are related to fundamental topics in genetics -- e.g. linkage disequilibrium (LD), the distribution of causal variants, and partitioning heritability -- which can lead to badly biased heritability estimates  \citep{zaitlen2012heritability,speed2012improved,gusev2013quantifying,gusev2014partitioning}.  

LMM-based heritability methods typically require specification of a genetic relationship matrix (GRM), which measures genetic similarity between subjects in a study.  The GRM may be based on familial or other information \citep{lange2003mathematical,powell2010reconciling}.  In GWAS, the GRM is frequently a sample correlation matrix constructed from study participant's trinary single nucleotide polymorphism (SNP) values, which corresponds to a Euclidean distance kernel \citep{yang2010common,zaitlen2012heritability}. In this paper, we argue that using the Euclidean kernel is a root cause for biases in LMM heritability estimation referenced in the previous paragraph.  Furthermore, we argue that if one adopts a Mahalanobis kernel-based GRM, then many of the LMM heritability biases related to LD and causal variants can be tranparently explained and resolved at the modeling level.  This approach yields a modified version of LMM-based heritability, relying on the Mahalanobis GRM. We also define a natural version of partitioned heritability with the Mahalanobis GRM, which resolves some closely related inconsistency issues that have been noted for other methods for partitioning heritability \citep{speed2012improved,gusev2013quantifying,gusev2014partitioning}. We propose a Mahalanobis kernel-based maximum likelihood estimator for both partitioned and total heritability, and show that the estimator is consistent and asymptotically normal.  Throughout the paper, numerical simulations are used to help illustrate different aspects of heritability estimation and our methodological results.

Beyond their immediate practical application for estimating heritability, the statistical arguments for the Mahalanobis kernel in this paper also address several fundamental questions about statistical modeling in modern genetics, including 1) fixed-effects vs. random-effects modeling and 2) narrow-sense vs. broad-sense heritability.  Questions about fixed- and random-effects modeling have been raised repeatedly in research on heritability estimation \citep{gibson2012rare}. Many of these questions can be summarized as follows: Should genetic effects be modeled as fixed or random quantities?  To answer this question, we argue that for the Mahalanobis kernel, the fixed- and random-effects models are essentially equivalent.  Furthermore, under the Mahalanobis kernel, we show that the LMM heritability coefficient can also be interpreted as a conditional variance -- which we refer to as the {\em $C$-heritability} ($C$ for ``conditional'') -- under the corresponding fixed-effects model.  This builds a link between narrow-sense (or additive) heritability, which LMM-based methods have traditionally been designed to estimate, and broad-sense heritability, which is a more model-free measure of overall heritability defined in terms of the conditional variance of a phenotype given the genotype and other specified information. By introducing $C$-heritability, total and partitioned heritabilities become special cases of a general form, and we are able to propose a unified approach for both total and partitioned heritability estimation.

The rest of this paper is organized as follows. In Section \ref{sec:lmm}, we review some of the common LMM heritability estimation methods and their model assumptions. Many of the biases and inconsistencies mentioned above are essentially related to model misspecification, which is discussed in detail in Section \ref{sec:challenges}. In Section \ref{sec:solutions}, we introduce the Mahalanobis distance-based approach, compare it with Euclidean GRM methods, and propose Mahalanobis-based estimators for heritability and partitioned heritability; $C$-heritability is also introduced and discussed in this section. Section \ref{sec:h2 simulations} contains additional simulation results and Section \ref{sec:discussion} contains a concluding discussion.   

\subsection{Related work}

Recently, in independent work, \cite{mathew2018novel} proposed using the Mahalanobis kernel in a similar way for heritability esitmation with GWAS data.   Mathew et al.'s paper primarily focuses on empirical analysis, using both simulated and real datasets to illsutrate advantages of the Mahalanobis kernel.  The present work contains more precise mathematical and statistical justification for much of the work in \cite{mathew2018novel}, and introduces statistical principles (e.g. $C$-heritability in Section \ref{sec:solutions}) that can be extended to other targeted application areas and genetics (like partitioning heritability). \section{LMMs for heritability estimation}\label{sec:lmm}

\subsection{Additive decomposition: From GRMs to LMMs}

In this section, we describe a statistical model that forms the basis for many LMM heritability methods for GWAS, \citep[e.g.][]{yang2010common,zaitlen2012heritability}.  Let $\y = (y_1,\ldots,y_n)^{\T} \in \R^n$ be a vector of centered, real-valued outcomes, where $y_i$ represents the phenotypic value of individual $i$ in some population.  Assume that 
\begin{equation}\label{additiveModel}
\y = \g + \boldsymbol{e}
\end{equation} can be decomposed as the sum of an additive genetic effect $\g = (g_1,\ldots,g_n)^{\T} \in \R^n$ and an uncorrelated noise vector $\boldsymbol{e} = (e_1,\ldots,e_n)^{\T} \in \R^n$, which may contain other non-additive genetic effects, environmental noise, and measurement error.  Further assume that  the data are centered, so that $\E(\g) = \E(\boldsymbol{e}) = 0$, and that $\Cov(\g) = \s^2_gK$ and $\Cov(\boldsymbol{e}) = \s^2_eI$, where $\s^2_g, \s^2_e \geq 0$ are genetic and environmental real-valued variance components, respectively, and $K$ is the $n \times n$ GRM. 

Thus, $\y$ is a random vector with $\E(\y)=0$ and $\Cov(\y)=\s^2_gK + \s^2_eI$, which we denote by
\begin{equation}\label{mvLmm}
\y \sim \mathcal{MV}(0,\s^2_gK + \s^2_eI).
\end{equation}
The heritability coefficient is defined to be
\begin{equation}\label{h2}
h^2 = \frac{\s^2_g}{\s_g^2 + \s_e^2}.
\end{equation}  
This definition of heritability is often referred to as narrow-sense or additive heritability.  
The GRM $K$ is typically normalized so that its diagonal entries all equal 1, so that the correlation matrix for $\y$ is $\mathrm{Corr}(\y) = h^2K + (1 - h^2)I$ and the heritability parameter $h^2$ represents the extent to which correlation between phenotypes in the population is determined by genetic relatedness.  

With GWAS data, genetic relatedness can be encoded by similarities between sequences of SNPs.  Let $\z_i = (z_{i1},\ldots,z_{im})^{\T}$ be the vector of normalized SNPs for the $i$-th study subject, i.e. 
\bes
z_{ij}=\frac{f_{ij}-2p_j}{\sqrt{2p_j(1-p_j)}},
\ees
where $f_{ij}=0,1,2$ is the minor allele count at SNP $j$ for individual $i$ and $p_j$ is the minor allele frequency (MAF) of SNP $j$ across the population  \citep{meuwissen2001prediction,hayes2009increased,zaitlen2012heritability} (in many studies, $m$ may be in the hundreds of thousands or millions). Then 
the $ij$-entry of the GRM $K = (K_{ij})$ is determined by some kernel function $K:\R^m \times \R^m \to \R$, whereby $K_{ij} = K(\z_i,\z_j)$. 

Traditionally, the GRM (also referred to as the kinship matrix) indicates the proportion of identical genetic regions that individual $i$ and $j$ inherited from common ancestors. This identity-by-descent (IBD) kernel is defined with respect to a pedigree, but knowledge of an explicit pedigree for the population in the study is usually infeasible in GWAS. In the absence of pedigree information, the GRM is frequently defined by the identity-by-state-based (IBS-based) GRM, where 
\begin{equation}\label{linearKernel}
K(\z_i,\z_j) = \frac{1}{m}\z_i^{\T}\z_j.
\end{equation}
The IBS-based GRM definition corresponds to the normalized Euclidean kernel; it measures average allelic correlations \citep{powell2010reconciling, speed2015relatedness} and is frequently used for GWAS. 
Other kernel functions have been proposed  for GWAS heritability estimation problems, e.g. the Gaussian kernel or higher-order polynomial kernels \citep{akdemir2015locally} and, recently, the Mahalanobis kernel \citep{mathew2018novel}, which is the focus of this paper. However, to date, the Euclidean kernel remains the most widely used and there is limited work in the literature on why one should prefer one GRM kernel over another.

The Euclidean kernel \eqref{linearKernel} corresponds to a linear random-effects model --- or a linear mixed model (LMM), if fixed-effects covariates are also included in the model --- hence, the term LMM-based heritability estimation.  For simplicity, in practice, the response value $\y$ is often onto the orthogonal complement of a subspace spanned by fixed-effects covariates such as sex, age, handedness, and leading eigenvectors of the genotype matrix \citep{visscher2008heritability,yang2011gcta,bonnet2015heritability,lee2016partitioning}. To see the correspondence between Euclidean kernel and linear random-effects model, let $\g = Z\u$ in \eqref{additiveModel}, where 
\begin{equation}\label{iidu}
\u = (u_1,\ldots,u_m)^{\T} \in \R^m, \ u_i \sim \mathcal{MV}(0,\s_g^2/m),
\end{equation}
is a vector of independent random genetic effects and $Z = (\z_1,\ldots,\z_n)^{\T}$ is the $n \times m$ matrix of genotypes.  Then \eqref{mvLmm} holds with the Euclidean kernel and we can rewrite \eqref{additiveModel} as
\begin{equation}\label{lmm}  
\y = Z\u + \boldsymbol{e}.
\end{equation}
In this model, the data from each subject is the (phenotype, genotype)-pair $(y_i,\z_i) \in \R^{m+1}$.

The main focus of this paper is the Mahalanobis kernel \citep{mahalanobis1936generalized,de2000mahalanobis}.  Let $\Sigma$ be the $m \times m$ positive definite matrix representing the population-level covariance (linkage disequilibrium) matrix for the SNPs $\z_i$, i.e. $\Sigma = \mathrm{Cov}(\z_i)$.  The Mahalanobis kernel is defined by 
\begin{equation}\label{mahalanobis}
K(\z_i,\z_j) = \z_i^{\T}\Sigma^{-1}\z_j;
\end{equation} 
it corresponds to a linear random-effects model with correlated random effects $\u\sim \mathcal {MV}(0,\tau^2_g/m\S^{-1})$ in \eqref{lmm}. The Mahalanobis kernel has been widely used in other applications involving genetics as a method to account for LD, e.g. genetic association testing \citep{majumdar2015semiparametric}. However, until recently \citep{mathew2018novel}, the Mahalanobis kernel has received less attention for heritability estimation.  

\subsection{Estimating $h^2$}\label{sec:h2est}
The method of moments and maximum likelihood are two widely used methods for estimating $h^2$ under \eqref{mvLmm}.  Both methods are discussed in this section and can be used for any GRM $K$.  In this section, we assume that the GRM is normalized with its diagonal entries all equal 1 and $\y$ is centered and that \eqref{mvLmm} holds.  

One of the classical moment estimators for $h^2$ comes from observing that $\sigma_g^2$ is the least squares regression coefficient for regressing $y_iy_j$ on $K_{ij}$ for all $i<j$. This is because \eqref{mvLmm} implies that
\bes
\E(y_iy_j|K)=\s_g^2 K_{ij}, \quad \mbox{for}\ i\neq j.
\ees
The corresponding estimator for $\s_g^2$ is
\bes
\tilde{\s}_g^2=\left(\widehat\Var(K_{ij})\right)^{-1}\widehat\Cov(y_i y_j,K_{ij}),
\ees
where 
\bes
\widehat\Var(K_{ij})&=&\frac{2}{n(n-1)}\sum_{i<j}K_{ij}^2,
\cr\widehat\Cov(y_iy_j,K_{ij})&=&\frac{2}{n(n-1)}\sum_{i<j}y_iy_jK_{ij}.
\ees 
\cite{henderson1984applications} used least squares in this way to estimate $h^2$ with
\bes
\tilde{h}^2=\frac{\tilde\s_g^2}{\|\y\|^2_2/n},
\ees
and variants of this method are still used today \citep{golan2014measuring,bulik2015ld,zhou2017unified,schwartzman2017simple}; this approach is also referred to as Haseman-Elston regression \citep{haseman1972investigation}.

To estimate $h^2$ using maximum likelihood, one typically assumes that $\y$ is Gaussian, i.e.
\[
\y \sim \mathcal N(0,\s_g^2K + \s_e^2I),
\]
and estimates $\s_g^2$, $\s_e^2$ and, subsequently, $h^2$, by maximizing the Gaussian likelihood for this model.  Specifically, let $\eta^2=\sigma^2_g/\sigma^2_e$.  This is a convenient reparametrization for the problem and $\eta^2$ can be interpreted as the signal-to-noise ratio. The maximum likelihood estimator for $(\sigma_e^2,\eta^2)$ is
\begin{equation}\label{vc}
(\hat{\sigma}^2_e,\hat{\eta}^2)=\argmax_{\sigma_e^2,\eta^2>0} l(\sigma_e^2,\eta^2),
\end{equation}
where
\bes l(\sigma^2_e,\eta^2)&=&-\frac{1}{2}\log(\sigma_e^2)-\frac{1}{2n}\log\det(\eta^2/m K +I)
\cr&&-\frac{1}{2n\sigma^2_e}\y^\T(\eta^2/m K+I)^{-1}\y.
\ees
Hence, the MLE of $h^2$ is 
\begin{equation}\label{mlh}
\hat h^2=\frac{\hat\eta^2}{\hat\eta^2+1}.
\end{equation}
\cites{yang2010common} groundbreaking work established the LMM approach for heritability estimation in GWAS with maximum likelihood.

Both maximum likelihood and moment estimators for $h^2$ have nice statistical properties (e.g. consistency).  In some circumstances, maximum likelihood estimators may have advantages over moment estimators in terms of efficiency (reduced variance).  On the other hand, moment estimators have been the subject of renewed interest recently because of potential advantages related to computation and data privacy (as many data only disclose summary GWAS statistics for the population \citep{finucane2015partitioning,zhou2017unified}).  

\subsection{Estimating partitioned heritability}\label{sec:partition}
Studies on partitioning heritability seek to identify the heritability $h_{\mathcal{S}}^2$, which is attributable to a subset of SNPs $\mathcal{S} \subseteq [m] := \{1,\dots,m\}$ \citep{gusev2014partitioning,finucane2015partitioning}. Usually the SNPs are partitioned by functional areas such as chromosomes, levels of MAF and functional annotations \citep{davis2013partitioning}. Partitioned heritability is also frequently estimated under LMM models.

In \citep{yang2011genome,kostem2013improving,gusev2014partitioning}, $\y$ is assumed to follow a LMM with two variance components
\bes
\y = Z_{\mathcal{S}}\u_{\mathcal{S}} + Z_{\mathcal{S}^c} \u_{\mathcal{S}^c} + \boldsymbol{e},
\ees
where
\begin{equation}\label{2vc}
u_j \sim \left\{
\begin{array}{ll} 
\mathcal{MV}\left(0,\frac{\s_{\mathcal{S}}^2}{\vert \mathcal{S}\vert }\right), & \mbox{if } j \in \mathcal{S}, \\
\mathcal{MV}\left(0,\frac{\s_{\mathcal{S}^c}^2}{m - \vert \mathcal{S}\vert} \right), & \mbox{if } j \notin \mathcal{S}. \end{array} \right.
\end{equation}
Under this model, the heritability due to $\mathcal{S}$ is defined as
\bes
h_{\mathcal{S}}^2 = \frac{\s_{\mathcal{S}}^2}{\s_{\mathcal{S}}^2 + \s_{\mathcal{S}^c}^2 + \s_e^2},
\ees
and it can be estimated using maximum likelihood after further assuming a Gaussian model for $u_j$ and $\boldsymbol{e}$ \citep{yang2011gcta,yang2011genome,davis2013partitioning,gusev2014partitioning}.

\section{Model misspecification and LMM heritability estimation}\label{sec:challenges}

LMM methods for estimating $h^2$ may give biased results when used in settings where the generative model for $\y$ differs from \eqref{mvLmm}, i.e. under model misspecification.  This has been noted repeatedly in the heritability literature \citep{zaitlen2012heritability}, and is important because many of the leading generative models from genetics for linking phenotypes $\y$ and SNP values $\z$ differ substantially from \eqref{mvLmm} \citep{barrett2009genome,stahl2010genome,gibson2012rare}.  In this section, we discuss model misspecification for LMMs in one of the most commonly used generative models for heritability in GWAS:  Causal loci models.  We show through a basic simulation example that the Euclidean kernel for LMM heritability estimation can give biased results uner the causal loci model; we also show numerically that the Mahalanobis kernel heritability estimator remains unbiased.  In the following sections, we explain in more detail why the Mahalanobis kernel methods for heritability estimation are more robust to model misspecification.

\subsection{Causal loci models}\label{sec:causal}

Many genetics models hypothesize a collection of causal loci (or causal variants), which are fixed locations along the genome, where the specific nucleotide combination impacts the phenotype  --- other, non-causal loci are assumed to have no direct impact on the phenotype \citep{pritchard2001rare}. In the context of the LMM \eqref{lmm}, this is frequently encoded by taking $\mathcal{A} \subseteq [m]$ to be the collection of causal loci and assuming:  
\begin{equation}\label{causal}
\begin{array}{l} u_j  \sim F \mbox{ are independent for } j \in \mathcal{A}, \\
u_j = 0 \mbox{ if } j \notin \mathcal{A},
\end{array}
\end{equation}
where $F$ is some probability distribution.  Some popoular models also allow the causal effect distribution to depend on $j$, e.g. $u_j \sim F_j$ for $j \in \mathcal{A}$ where $F_j$ could depend on the MAF for the $j$-th SNP \citep{yang2015genetic,gazal2017linkage,schoech2017quantification}.

If $\mathcal{A} \neq [m]$, then the genetic effects assumption \eqref{causal} violates \eqref{iidu}.  Computing the expected value of the score equations for the variance components MLE \eqref{vc} under \eqref{causal} indicates that this alone may not be enough to induce bias in heritability estimates, i.e. the score equations may remain unbiased in some cases.  However, if the SNPs are in linkage disequilibrium (i.e. $\Cov(\z_i)$ is not diagonal) and if the causal SNPs are not uniformly distributed across the genotyped SNPs, then heritability estimates are frequently badly biased. This has been noted previously in the literature, e.g. \cite{zaitlen2012heritability,speed2012improved,gusev2013quantifying,yang2015genetic}.

\subsection{Model misspecification in partitioned heritability estimation}
Similarly, in partitioned heritability estimation, care must be taken when disentangling the effects of SNPs in $\mathcal{S}$ with SNPs that are in linkage disequilibrium with $\mathcal{S}$. 

In particular, if LD is ignored, then estimates of $h^2_{\mathcal{S}}$ can be biased, as common generative models for paritioned heritability typically differ from \eqref{2vc}. Under the causal loci hypothesis, individual effect-size follows
\bes
u_j \left\{
\begin{array}{ll} 
	\sim F_{S} &\mbox{are independent for } j \in \mathcal{A}_1\subseteq \mathcal{S},\\
	\sim F_{S^c} &\mbox{are independent for } j \in \mathcal{A}_2\subseteq{S}^c,\\
	=0&\mbox{otherwise,}\end{array} \right.
\ees
where $\mathcal{A}_1$ and $\mathcal{A}_2$ are the sets of causal loci in $\mathcal{S}$ and $\mathcal{S}^c$, with $\mathcal{A}_1 \cap \mathcal{A}_2=\emptyset$ \citep{gusev2014partitioning}. 
If the causal loci are concentrated in an uneven LD region, then similar bias observed in total heritability estimation is expected for partitioned heritability estimation. Simulation results illustrating this bias are discussed in Section \ref{sec:simulations.partitioned}, after introducing the Mahalanobis estimator for partitioned heritability.

\subsection{Other existing methods for improving heritability estimates}
Many strategies have been proposed to account for potential bias in LMM-based heritability estimates.  
One simple strategy to improve LMM heritability estimates is LD pruning:  One SNP from each pair of highly correlated SNPs is simply removed from the analysis \citep{purcell2007plink,stahl2012bayesian}.  A drawback of this approach is that without information about $\mathcal S$, causal loci could potentially be removed during the pruning step, which may induce additional biases when estimating $h^2$. 
Other strategies focus on transforming and re-weighting the genotype matrix $Z$. For example \cite{gusev2013quantifying} built on work of \cite{patterson2006population}, and proposed to transform the genotype matrix by regressing each SNP on all preceeding SNPs. Each SNP genotype is then replaced by the regression residuals. Similarly, the LD adjusted kinship (LDAK)  method suggests assigning different weights to SNPs  \citep{speed2012improved}.  Optimal SNP weights are computed by considering local LD and distance to neighboring SNPs, then solving a linear programming problem.  The reweighted data is then analyzed with LMM methods. 

Many of these bias-correction methods for mitigating the impact of LD on heritability estimation can be intepreted as modifying the kernel matrix $K$ in \eqref{vc}; however, modifying the kernel is not typically their primary motivation.  The general solution proposed in this paper is to replace the Euclidean kernel typically used for LMM-based heritability estimation with the Mahalanobis kernel.  The next subsection contains a simulation example, which shows that the Gaussian MLE for $h^2$ can be biased under the Euclidean kernel, but that estimators with the Mahalanobis kernel are unbiased.  Methodological and theoretical justification for the Mahalanobis kernel is provided in Section \ref{sec:solutions}.  A more detailed simulation study is contained in Section \ref{sec:simulations.full}, where some of the other methods for mitigating bias in heritability estimation mentioned in this section (e.g. LDAK) are also considered.  

\subsection{Simulation example}

In the simulations considered in this section, we assume the model \eqref{causal} holds, with Gaussian $F$.  Let $Z_{\mathcal{A}}$ denote the $n \times \vert \mathcal{A}\vert$ matrix obtained by extracting the columns of $Z$ corresponding to $\mathcal{A}$.  If $e_j$ are Gaussian and $u_j \sim \mathcal{N}(0,\s_g^2/\vert \mathcal{A} \vert)$ are iid Gaussian causal effects, for $j \in \mathcal{A}$, then
\[
\y \sim \mathcal{MV}\left(0,\frac{\s_g^2}{\vert \mathcal{A}\vert}Z_{\mathcal{A}}Z_{\mathcal{A}}^{\T} + \s^2_eI\right).
\]
Thus, $\y$ follows the model \eqref{mvLmm} with $K_{i,j} = K(\z_i,\z_j) = \z_{i,\mathcal{A}}^{\T}\z_{j,\mathcal{A}}/\vert \mathcal{A}\vert$ and $\z_{i,\mathcal{A}} = (z_{ik})_{k \in \mathcal{A}} \in\R^{|\mathcal{A}|}$, so the heritability coefficient is $h^2 = \s_g^2/(\s_g^2 + \s_e^2)$.  On the other hand, in the absence of additional information about $\mathcal{A}$, LMM heritability estimators $h^2$ are typically fit according to the model \eqref{lmm}, with the Euclidean kernel \eqref{linearKernel}.   In this simulation study, we estimated $h^2$ under this setting using the MLE \eqref{vc} with the Euclidean kernel \eqref{linearKernel} and the MLE with the Mahalanobis kernel \eqref{mahalanobis}.

In our simulation study, we took:
\begin{itemize}
	\item[(i)] $n = 500$, $m = 1000$.
	\item[(ii)] $\mathcal{A} = \{1,\ldots,m/2\}$.
	\item[(iii)] $ \sigma_g^2 = \sigma_e^2 = 0.5$.
	\item[(iv)] $\z_1,\ldots,\z_n \sim \mathcal N(0,\S)$, where
	\[
	\S = \left(\begin{array}{cc} \mathrm{AR}(0.3) & 0 \\ 0 & \mathrm{AR}(0.7) \end{array}\right)
	\]
	and $\mathrm{AR}(\rho)$ is the $m/2 \times m/2$ matrix with $ij$-entry $\rho^{\vert i - j\vert}$.  
\end{itemize}
Under this setup, $h^2 = 0.5$.  We simulated 50 independent datasets specified according to this model, and computed the Euclidean and Mahalanobis kernel MLEs for each dataset.
Summary statistics are reported in Table \ref{sim1}.
\begin{table}[h]
	\begin{center}
		\caption{Means and confidence intervals for estimates of $h^2$. Based on results from 50 independent datasets. $h^2$ is estimated by MLE with linear and Mahalanobis kernels.}
		\label{sim1}
		\vspace{.1in}
		\begin{tabular}{c|lc|lc}
			$h^2$ &  \multicolumn{2}{|c|}{Linear MLE} &  \multicolumn{2}{c}{Mahalanobis MLE} \\ \hline
			0.5 &   Mean: &0.454 & Mean: & 0.495 \\
			&  95\% CI:  & (0.427,0.482)& 95\% CI: & (0.468, 0.522)
		\end{tabular}
	\end{center}
\end{table}
From Table \ref{sim1}, it's evident that the estimator based on the Euclidean kernel is significantly biased, and the Mahalanobis estimator is not.

\section{Fixed-effects models and $C$-heritability -- Why the Mahalanobis kernel works}\label{sec:solutions}
\subsection{Fixed-effects heritability}\label{sec:FE}

The potential impact of model misspecification on random-effects heritability estimation was illustrated in the previous section.  Our approach to resolving this problem begins by treating the genetic effects as fixed, rather than random quantities.  The fixed-effects approach has been discussed elsewhere in the literature \citep[e.g.][]{price2010new}, but our focus on the Mahalanobis kernel appears to be new.  One of the main arguments in previous literature for fixed genetic effects is interpretability: In reality, the effect of a given SNP on a phenotype is fixed and statistical analysis should be conducted conditional on these effects.  In this section, we show that starting from a fixed-effects model with random genotype, we can recover the Mahalanobis kernel LMM estimator and prove that it has nice (asymptotic) statistical properties for heritability estimation in both fixed-effects model {\em and} random-effects models with nearly arbitrary genetic effect distribution.  

Assume that the linear model \eqref{lmm} holds with some fixed (non-random) $\u$.  Additionally assume that
\begin{equation}\label{normality}
\z_1,\ldots,\z_n \sim \mathcal N(0,\Sigma) \mbox{ and } e_1,\ldots, e_n \sim \mathcal N(0,\s^2_e)
\end{equation}
are independent. Thus, this is a random design (or random genotype) model, as opposed to a random-effects model.  The random design assumption is important for our analysis; however, normality is probably not essential.  Indeed, the normality assumptions are unrealistic in practice (the entries of $\z_i$ are typically discrete).  Primarily, we rely on the normality assumption for motivating the methods proposed in this section.  Many other high-dimensional variance component estimation with fixed-effects model \citep[e.g.][]{dicker2014variance,janson2017eigenprism} require the same multivariate Gaussian random-design \eqref{normality}, for its invariance property under orthogonal transformations. Work of \cite{bai2007asymptotics} in random matrix theory has shown that in the large limit where $n,m\to\infty$, the invariance property holds for a broader class of random matrices.  We expect our estimator to be robust asymptotically for reasonable random designs, as suggested by simulation results in Section \ref{sec:h2 simulations} and related numerical results in \citep{janson2017eigenprism}. Theoretical results on relaxing the Gaussian random design assumptions for the Mahalanobis estimator (e.g. by building on results in \citep{dicker2016flexible,dicker2016maximum}) would be an interesting future research direction. 

Let $(y,\z)$ be a generic draw from the study population.  We define the fixed-effects heritability coefficient to be
\begin{equation}\label{FE}
h^2 = 1 - \frac{\E\left(\Var(y\mid \z)\right)}{\Var(y)} = \frac{\u^{\T}\S\u}{\u^{\T}\S\u + \s_e^2}.  
\end{equation}
This is a version of broad-sense heritability, determined by the conditional variance of the phenotype \citep{visscher2008heritability}.   The fixed-effects heritability \eqref{FE} captures correlation between SNPs and LD through the quadratic form $\u^{\T}\S\u$.  

\subsection{Fixed- vs. random-effects heritability}\label{sec:RFE}

In general, the fixed-effects heritability coefficient \eqref{FE} differs from the random-effects heritability coefficient \eqref{h2} under the Euclidean kernel.   Indeed, let $h^2_{FE}$ denote the fixed-effects heritability coefficient \eqref{FE}, and let $h^2_{Euc}$ denote the random-effects heritability coefficients corresponding to the model \eqref{mvLmm} with Euclidean kernel.  Without loss of generality, we assume that the data are normalized so that $\Var(y)=1$. Then \eqref{FE} becomes $h^2 = h^2_{FE}=\Var(\z^\T\u|\u)=\u^\T\S\u$.   If $\S=I$, note that
 \begin{equation}
 \label{euc}
 h^2_{FE}=\|\u\|^2_2.
 \end{equation} 
 Under the random-effects heritability model corresponding to the Euclidean kernel, $\u \sim \mathcal{MV}(0,\s^2_g/mI)$ and the heritability coefficient is $h_{Euc}^2 = \s_g^2$ (recall that $\Var(y) = 1$); thus, $h^2_{FE} = \|\u\|^2_2 \approx \s^2_g = h^2_{Euc}$ when $m$ is large.  On the other hand, if $\S \neq I$ (i.e. if there is LD), then \eqref{euc} does not hold and $h^2_{FE}$ may differ substantially from $h^2_{Euc}$.  

Next, we show that the fixed-effects heritability coefficient is asymptotically equivalent to the random-effects heritability coefficient under the model \eqref{mvLmm} with Mahalanobis kernel -- this is the key argument for the Mahalanobis kernel methods in this paper.  Consider the random-effects model \eqref{mvLmm} under the Mahalanobis kernel \eqref{mahalanobis}, where $\u \sim \mathcal{MV}(0,h^2_{FE}/m\S^{-1})$. Under this model, the random-effects heritability coefficient is
\bes
h^2 = h^2_{Mah}=\Var(\z^\T\u|\z)=\Var(\z^\T \S^{-1/2}\S^{1/2}\u|\z)=\z^\T\S^{-1}\z\E(\u^{\T}\S\u) \approx h^2_{FE}\z^{\T}\S^{-1}\z/m \approx h^2_{FE},  
\ees
where the approximation is valid for large $m$.
Hence, random-effects heritability under the Mahalanobis kernel is approximately equivalent to fixed-effects heritability.  Moreover, the fixed-effects approach -- and, hence, the Mahalanobis approach -- does not depend on the distribution of causal loci, which can be a source of bias for Euclidean kernel methods.

\subsection{Partitioning heritability}\label{sec:partitioning heritability}

Broad-sense heritability and the fixed-effects linear model described in Section \ref{sec:FE} also motivate a natural definition of partitioned heritability.  For $\mathcal{S} \subseteq [m]$, we define the heritability attributable to $\mathcal{S}$ to be  
\begin{equation}\label{fep}
h^2_{\mathcal{S}} = 1 - \frac{\E\left(\Var(y \mid \z_{\mathcal{S}})\right)}{\Var(y)} = \frac{\u^{\T}\S\u - \u^{\T}_{\mathcal{S}^{c}} \S_{\mathcal{S}^c\mid \mathcal{S}}\u_{\mathcal{S}^{c}}}{\u^{\T}\S\u + \s_e^2},
\end{equation}
where $\S_{\mathcal{S}^c\mid \mathcal{S}} = \Sigma_{\mathcal{S}^c,\mathcal{S}^c} - \Sigma_{\mathcal{S}^c,\mathcal{S}}\S_{\mathcal{S},\mathcal{S}}^{-1}\Sigma_{\mathcal{S},\mathcal{S}^c}$ and $\S_{\mathcal{S}_1,\mathcal{S}_2}$ is the submatrix of $\S$ with rows and columns selected according to $\mathcal{S}_1,\mathcal{S}_2 \subseteq [m]$, respectively.  This definition for partitioned heritability consistently accounts for correlation between LD and SNPs.  Additionally, note that $h^2 = h^2_{[m]}$.  

The definition \eqref{fep} makes sense in the context of broad-sense heritability, and when the genetic features are Gaussian (or approximately Gaussian).  However, as discussed in Section \ref{sec:FE}, the Gaussian assumption basically never holds in practice.  In the following proposition, we argue that the definition \eqref{fep} is also a natural consequence of three reasonable properties we might expect of any quadratic form-based estimator for partitioned heritabilty for linear models with fixed genetic effects.  

\begin{prop}\label{prop:ph}
	Assume that the linear model \eqref{lmm} holds, that $\u \in \R^m$ is a fixed vector and that $\E(\z) = 0$, $\Var(\z) = \S$.  Assume that the heritability attributable to $\mathcal{S}$, $h^2_{\mathcal{S}} = h^2_{\mathcal{S}}(\u; \S)$, is a quadratic form in $\u$, and that $h^2_{\mathcal{S}}$ satisfies the following properties:
	\begin{itemize}
		\item[(i)] $0\leq h^2_{\mathcal{S}}(\u;\S)\leq h^2(\u;\S)$ for all $\u \in \R^m$, where $h^2 =h^2(\u;\S)$ is the fixed-effects heritability \eqref{FE},\label{ph:property1}
		\item[(ii)] $h^2_{\mathcal{S}}(\u;\S) =h^2(\u;\S)$ if and only if $\u_{\mathcal{S}^c}=0$, and \label{ph:property2}
		\item[(iii)] $h^2_{\mathcal{S}}(
		\u; \S)$ does not depend on $\S_{\mathcal{S}^c,\mathcal{S}^c}$\label{ph:property3}.
	\end{itemize}
	Then we must have
	\[
	h^2_{\mathcal{S}} = \frac{\u^{\T}\S\u - \u^{\T}_{\mathcal{S}^{c}} \S_{\mathcal{S}^c\mid \mathcal{S}}\u_{\mathcal{S}^{c}}}{\u^{\T}\S\u + \s_e^2}.
	\]
\end{prop}

Proposition \ref{prop:ph} is proved in the Appendix.  Condition (i) in Proposition \ref{prop:ph} says that the heritability attributable to a subset of SNPs $\mathcal{S}$ must be smaller than the total heritability (i.e. the heritability attributable to all measured SNPs); condition (ii) means that the heritability attributable to $\mathcal{S}$ is equal to the total heritability if and only if all causal loci are contained in $\mathcal{S}$; condition (iii) means that the heritability attributable to $\mathcal{S}$ should not depend on LD amongst SNPs that are not in $\mathcal{S}$ (though it 
may depend on LD between SNPs in $\mathcal{S}$ and those not in $\mathcal{S}$). We'll discuss how to estimate $h^2_{\mathcal{S}}$ in Section \ref{sec:estimation C}.

\subsection{$C$-heritability with projections}\label{sec:proj}

In addition to focusing on the heritability attributable to a subset of SNPs $\mathcal{S}$ with partitioned heritability, we can extend the definition of heritability to variation explained by any linear projection $C^{\T}\z$, for $m \times k$ matrices $C$ with rank $k$:
\begin{equation}\label{C-broad}
h^2_C = h^2_C(\u; \S) = 1 - \frac{\E\left(\Var(y \mid C^{\T}\z)\right)}{\Var(y)}.
\end{equation}
$C$-heritability can be used to describe both total heritability (where $C = I$ is the identity matrix) and partitioned heritability (where $C$ is a coordinate projection matrix corresponding to $\mathcal{S}$).  More importantly, it is convenient to describe a generic method for estimating $C$-heritability, which is applicable to both total heritability and partitioned heritability estimation (in the former case, we will see that this is equivalent to Mahalanobis kernel-based methods).

Under the linear model with Gaussian data \eqref{normality} and the additional assumption that $\Var(y) = 1$, the C-heritability coefficient is given by
\bes
h^2_C=\u^\T \S C(C^\T \S C)^{-1}C^\T \S\u.
\ees
The following lemma summarizes some useful facts about $h^2_C$.  

\begin{lemma}\label{lemma:cher}
	Assume \eqref{lmm} and \eqref{normality} and that $\Var(y) = 1$.  Then $h^2_C(\u; I) = \u^{\T}C(C^{\T}C)^{-1}C^{\T}\u$ 
	and
	\begin{equation}\label{cher2}
	h^2_C(\u; \S) = h^2_{\S^{1/2}C}(\S^{1/2}\u;I).
	\end{equation}
	If, furthermore, $m = k$, then 
	\begin{equation}\label{cher3}
	h^2_C(\u; \S) = \u^{\T}\S\u = h^2.  
	\end{equation}
	Finally, let $\mathcal{S} \subseteq [m]$ and let $\Pi_{\mathcal{S}}$ be the projection matrix onto coordinates indexed by $\mathcal{S}$.  Then 
	\begin{equation}\label{cher4}
	h^2_{\mathcal{S}} = h^2_{\Pi_{\mathcal{S}}}.
	\end{equation}
\end{lemma}

The proof of Lemma \ref{lemma:cher} is trivial. The identity \eqref{cher2} helps to explain the connection between LD and heritability -- it implies that heritability in a model with LD structure $\S$ is equivalent to heritability in a model where LD has been removed through a whitening transformation $\z \mapsto \S^{-1/2}\z$.  The equation \eqref{cher3} implies that $C$-heritability is invariant under any (full rank) change-of-basis for the genotype $\z \mapsto C^{-1}\z$.  The last identity \eqref{cher4} shows how to estimate partitioned heritability $h^2_{\mathcal{S}}$, when combined with the results for estimating $C$-heritability in the next subsection.

\subsection{Estimating $C$-heritability}

\label{sec:estimation C}
In this subsection, we assume the fixed-effects linear model \eqref{lmm} with $\z\sim \mathcal{N}(0,\S)$. We would like to estimate $h^2_C(\u;\S)$, for a full rank matrix $C\in \mathbb{R}^{m\times k}$.  When $k = m$, we saw in Lemma \ref{lemma:cher} that $C$-heritability equals the total heritability, i.e. $h^2_C(\y;\S) = h^2$.  Moreover, the results in Section \ref{sec:RFE} show that total heritability in the fixed-effects model can be estimated in the same way as the LMM heritability coefficient with Mahalanobis kernel, e.g. we can use one of the methods described in Section \ref{sec:h2est} with $K$ given by \eqref{mahalanobis}. For projections $C$ with $k < m$, our strategy is to reduce the problem to a total heritability estimation problem with $k$ SNPs and use the methods just described, based on the Mahalanobis kernel/fixed-effects model equivalence.  

Assume that $k < m$ and let $U_C$ be a $m \times k$
matrix with orthonormal columns such that $\S^{1/2}C(C^{\T}\S
C)^{-1}C^{\T}\S^{1/2} = U_CU_C^{\T}$.  Let $U_{C^{\perp}}$ be a
corresponding $m \times (m-k)$ matrix with orthonormal columns
satisyfing $U_C^{\T}U_{C^{\perp}} = 0$ and $I = U_CU_C^{\T} +
U_{C^{\perp}}U_{C^{\perp}}^{\T}$.  Then
\begin{align}\nonumber
\y & = Z\u + \boldsymbol{e} \\ \nonumber
& =Z\S^{-1/2} U_CU_C^{\T} \S^{1/2}\u + Z\S^{-1/2}
U_{C^{\perp}}U_{C^{\perp}}^{\T}\S^{1/2} \u + \boldsymbol{e}\\ \nonumber 
& = W_C\v_C + W_{C^{\perp}}\v_{C^{\perp}} + \boldsymbol{e} \\ \label{zlm}
& = W_C\v_C + \boldsymbol{e}_C,
\end{align}
where
\begin{align*}
W_C   &= Z\S^{-1/2}U_C=Z C(C^\T \S C)^{-1/2},   \ \ 
W_{C^{\perp}}  = Z\S^{-1/2}U_{C^{\perp}}, \\
\v_C  &= U_C^{\T}\S^{1/2}\u,  \ \  \v_{C^{\perp}} =
U_{C^{\perp}}^{\T}\S^{1/2}\u, \\  \boldsymbol{e}_C &=
W_{C^{\perp}}\v_{C^{\perp}} + \boldsymbol{e}.
\end{align*}
Since $\boldsymbol{e}_C$ is independent of $W_C$, we've transformed the original linear model with
data $(\y,Z)$ into the linear model \eqref{zlm} with data $(\y,W_C)$, where 
\bes
(W_C)_{ij}&\sim& \mathcal{N}(0,1),\ \ 1\leq i\leq n, \  1\leq j\leq k \mbox{ are iid and}
\cr\boldsymbol{e}_C&\sim& \mathcal{N}\left(0,(\Vert \v_{C^\perp}\Vert^2+\sigma_e^2)I\right).
\ees
Moreover, since $\Vert \v_C\Vert^2 = \u^{\T}\S^{1/2}U_CU_C^{\T}\S^{1/2}\u = \u^{\T}\S C(C^{\T}\S
C)^{-1}C^{\T}\S\u = h^2_C$ when $\Var(y) = 1$, it follows that the total (fixed-effects) heritability $h^2$ for the model \eqref{zlm} is equivalent to the $C$-heritability $h^2(\u;\S)$ for the original linear model.  Thus, to estimate $h^2(\u;\S)$, we simply esimate the fixed-effects total heritability coefficient $h^2$ under \eqref{zlm}.  This is formalized in the following  proposition.  
\begin{prop}\label{ch-mle}
Assume that \eqref{lmm} and \eqref{normality} hold, and let $C$ be a full rank $m \times k$ matrix with $k \leq m$. Define $\s^2_{C^{\perp}} = \u^{\T}\S^{1/2}U_{C^{\perp}}U_{C^{\perp}}^{\T}\S^{1/2}\u + \s_e^2$ and $\eta_C^2 =  \u^{\T}\S^{1/2}U_CU_C^{\T}\S^{1/2}\u/\s^2_{C^{\perp}}$.  Additionally, let $W_C$ be as defined in \eqref{zlm} and let
\bes
\hat{h}^2_C=\frac{\hat{\eta}^2_C}{1+\hat{\eta}^2_C},
\ees
where 
\begin{align*}
(\hat{\eta}^2_C,\hat\sigma^2_{C^\perp}) & :=\argmax_{\eta^2_C,\ \sigma^2_{C^\perp}} - \frac{1}{2}\log(\sigma^2_{C^\perp})-\frac{1}{2n}\log \det
\left(\frac{\eta_C^2}{k} W_CW_C^{\T} + I\right)  \\ & \qquad
- 
\frac{1}{2n\sigma^2_{C^\perp}}\y^{\T}\left(\frac{\eta_C^2}{k}W_CW_C^{\T} +
I\right)^{-1}\y .
\end{align*}
If $n \to \infty$, $k/n \to \rho \in (0,\infty)\setminus\{1\}$, and $\s_{C^{\perp}}^2,\eta_C^2$ are both contained in some compact subset of $(0,\infty)$, then 
\begin{equation}\label{consistency}
\hat{h}^2_C \to h_C^2
\end{equation}
in probability, where $h_C^2$ is the $C$-heritability \eqref{C-broad}. Finally, define $\mathcal I=\frac{1}{k}W_CW_C^\T$ and $\mathcal J=\frac{\eta^2_C}{k}W_CW_C^\T+I$, then 
\begin{equation}\label{asympnorm}
	\sqrt{n}(\hat{h}^2_C-h^2_C)\overset{\mathcal D}{\longrightarrow}\mathcal{N}\left(0 ,\quad \frac{2\sigma^4_{C^\perp}}{(1+\eta^2_C)^4}\left(1-\frac{\tr(\mathcal I\mathcal J^{-1})^2}{n\tr(\mathcal{I}^2\mathcal{J}^{-2})}\right)^{-1}\right)^{-1}
\end{equation}
under the same asymptotic setting as \eqref{consistency}.  
\end{prop}

Proposition \ref{ch-mle} is proved in the Appendix.  Equations \eqref{consistency} and \eqref{asympnorm} imply that $\hat{h}^2_C$ is consistent and asymptotically normal.  Proposition \ref{ch-mle} is essentially a corollary of Theorems 1--2 in \citep{dicker2016maximum}.  Observe that when $C = I$ is the $m \times m$ identity matrix, $h^2_C = h^2$ is the fixed-effects total heritability coefficient and the estimator $\hat{h}_C^2$ in Proposition \eqref{ch-mle} is identically equal to the MLE \eqref{mlh} with the Mahalanobis kernel \eqref{mahalanobis}.

\section{Numerical experiments} \label{sec:h2 simulations} 

In this section we report on results from several numerical experiments on total and partitioned heritability estimation.  Data was simulated based on publicly available minor allele frequency data \footnote{Available at https://www.sanger.ac.uk/resources/downloads/human/hapmap3.html as of January 8, 2019.} from the third phase of the International HapMap Project (HapMap3) \citep{international2010integrating}. The study participants are Utah residents with European ancestry (CEU). Performance of total heritability estimators are compared while varying LD-level and sparsity of causal variants. Performance of partitioned heritability estimators are compared with different distributions of causal variants across the genetic-effect vector -- within and without the subset of interest -- and in relation to the LD structure. 

We consider the linear model
\bel{lmm.sim}
\y=Z\u+\e
\eel
where 
\begin{itemize}
	\item [(i)] $n=1,000$, $m=10,000$
	\item [(ii)]$\sigma^2_e=1-\sigma^2_g$, $0 < \s_g^2 < 1$
	\item [(iii)]$\e\sim\mathcal{N}(0,\sigma^2_e I) $
	\item [(iv)] 
	\bes
	z_{ij}=\frac{f_{ij}-2p_j}{\sqrt{2p_j(1-p_j)}},
	\ees
	where $f_{ij}\sim Binomial (2,p_j)$, and $p_j$ is the empirical MAF of the $j^{th}$ SNP in chromesome 1 of HapMap3 CEU samples \citep{international2010integrating}, where SNPs with $p_j\geq 5\%$ are targeted by the HapMap study. In addition to this, we consider the case that absolute difference between MAFs of adjacent SNPs less than $5\%$ in order to construct variables with a covariance structure $\S$ defined as follows.
	\item [(v)] Let $m_b=100$ and $\nu=m/m_b$. Let $\mathrm{AR}(\rho)$ be the $m_b \times m_b$ matrix with $ij$-entry $\rho^{\vert i - j\vert}$. Then $\Cov(\z_i)=\S$, where
	\[
	\S = \left(\begin{array}{cccc} D_1 & 0 &\dots &0 \\ 
	0& D_2&\dots&0 \\
	0&0&\ddots&0\\
	0&\dots&0&D_{\nu}\end{array}\right)
	\]
	and 
	$D_k=\begin{cases}
	AR(0.4), \ \text{if} \ k\leq \nu/2;\\
	AR(0.6), \ \text{otherwise.} 
	\end{cases}$ 
	\end{itemize}
	The multivariate random variable $\z_i$ with covariance $\S$ and marginal binomial distribution is simulated through the Gaussian copula method with its intermediate Gaussian correlation matrix recovered iteratively \citep{ferrari2012simulating,barbiero2017r}. The conditions on MAFs between adjacent SNPs in (iv) ensures that the simulation procedure is feasible for constructing $\S$ via the Gaussian copula. 
\subsection{Total heritability estimation}\label{sec:simulations.full} 
We estimated total heritability in the linear model \eqref{lmm.sim} with the simulation setup described above and genetic variances $\sigma^2_g=0.3,0.5,0.7$.  
We also considered different configurations of causal variants.  In particular, for each simulated dataset $d$, the genetic effects were simulated as follows, \bes
u_j^{(d)}\begin{cases}
	\sim \mathcal{N}(0,\psi_j),& \text{if} \ j\in \mathcal{A},\\
	=0,& \text{otherwise,}
\end{cases}
\ees
where $\psi_j=\frac{1}{c}\sigma^2_g (p_j(1-p_j))^{-1} $ and $c$ is a normalizing constant which ensures that $\sum_{j \in A} \psi_j = \s_g^2$ \citep{speed2012improved,lee2013estimation}. 

\begin{figure}[H]
	\begin{center}
		\includegraphics[width=0.495\textwidth]{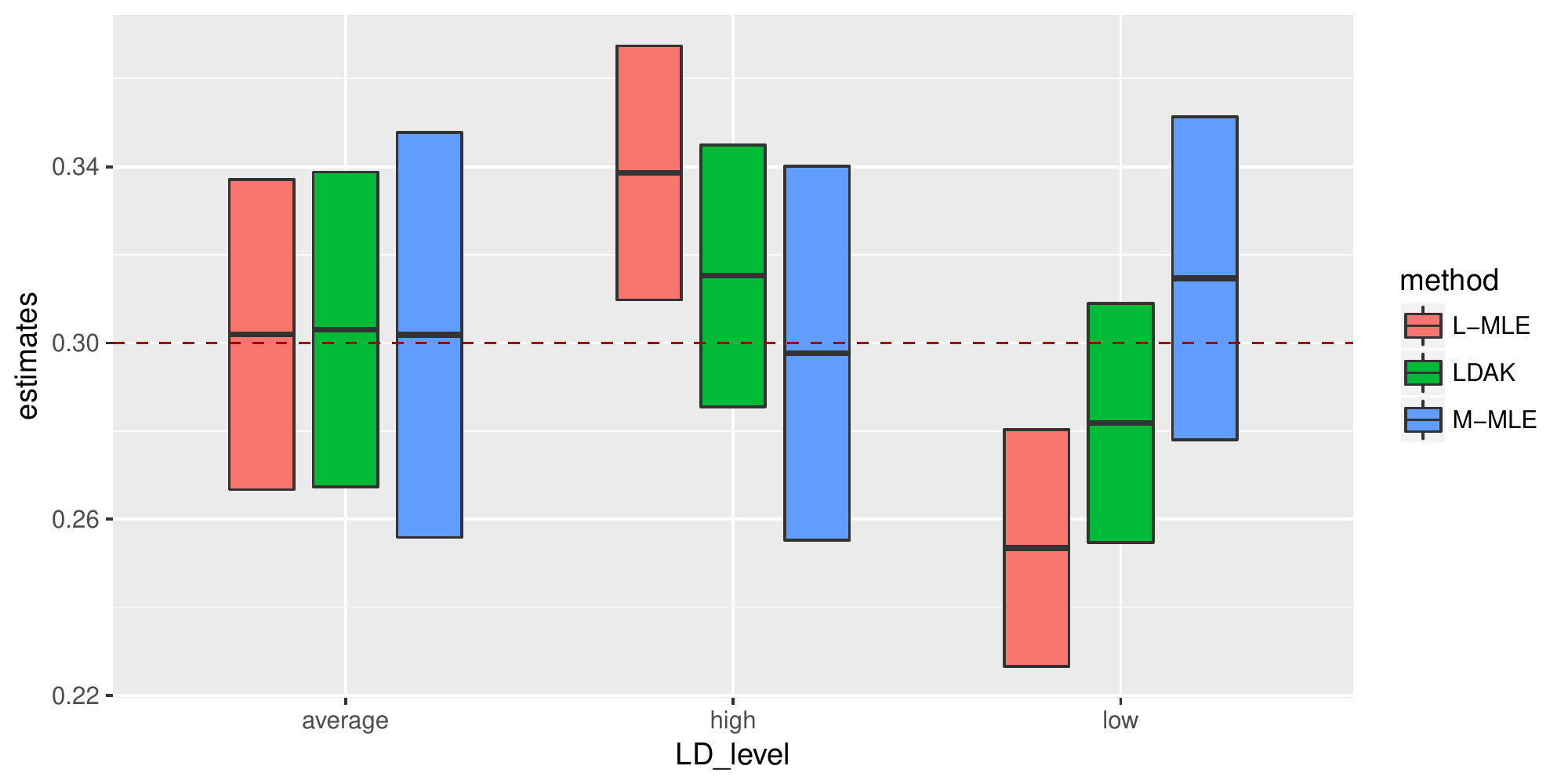}
		\includegraphics[width=0.495\textwidth]{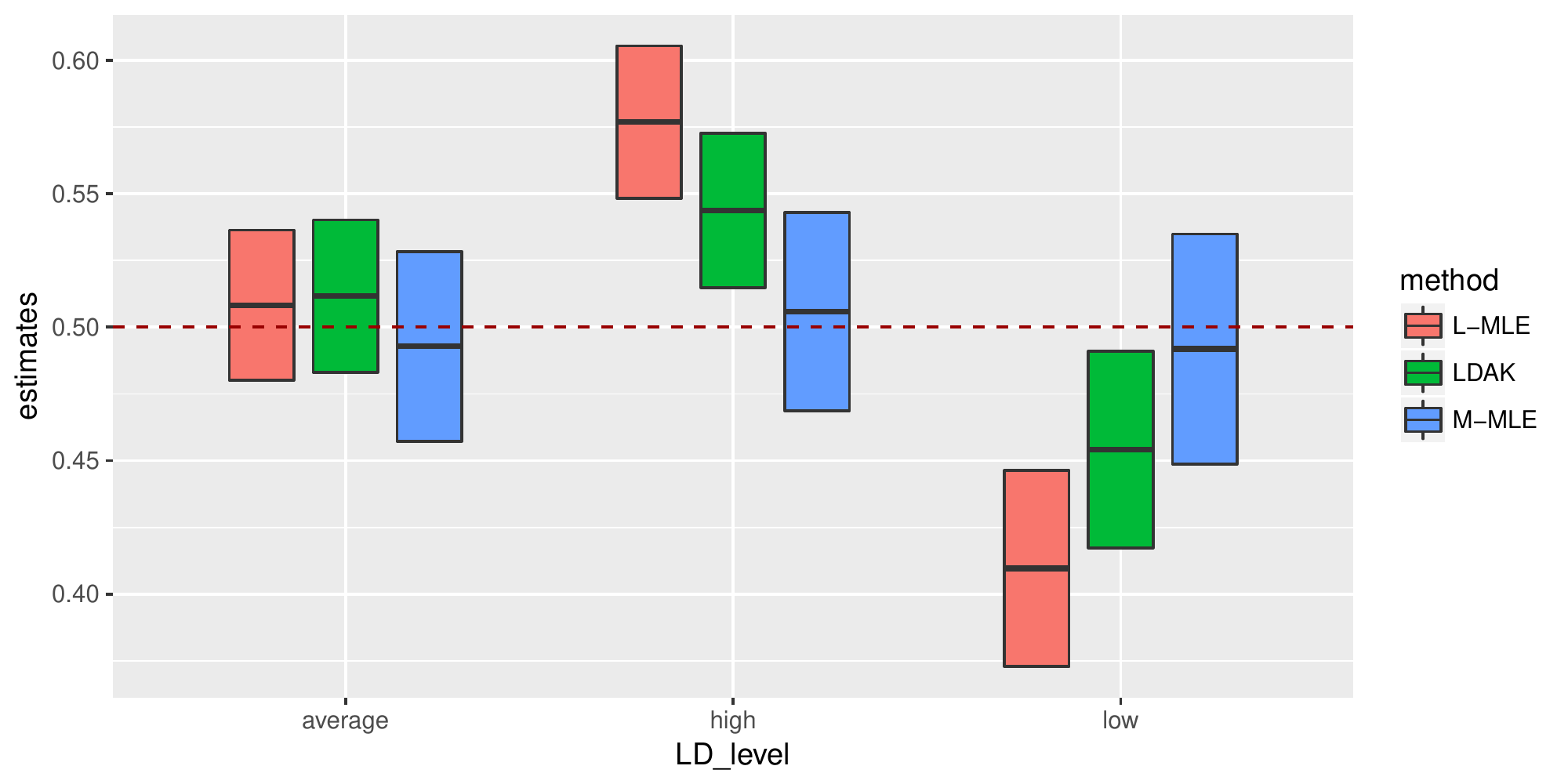}
		\includegraphics[width=0.495\textwidth]{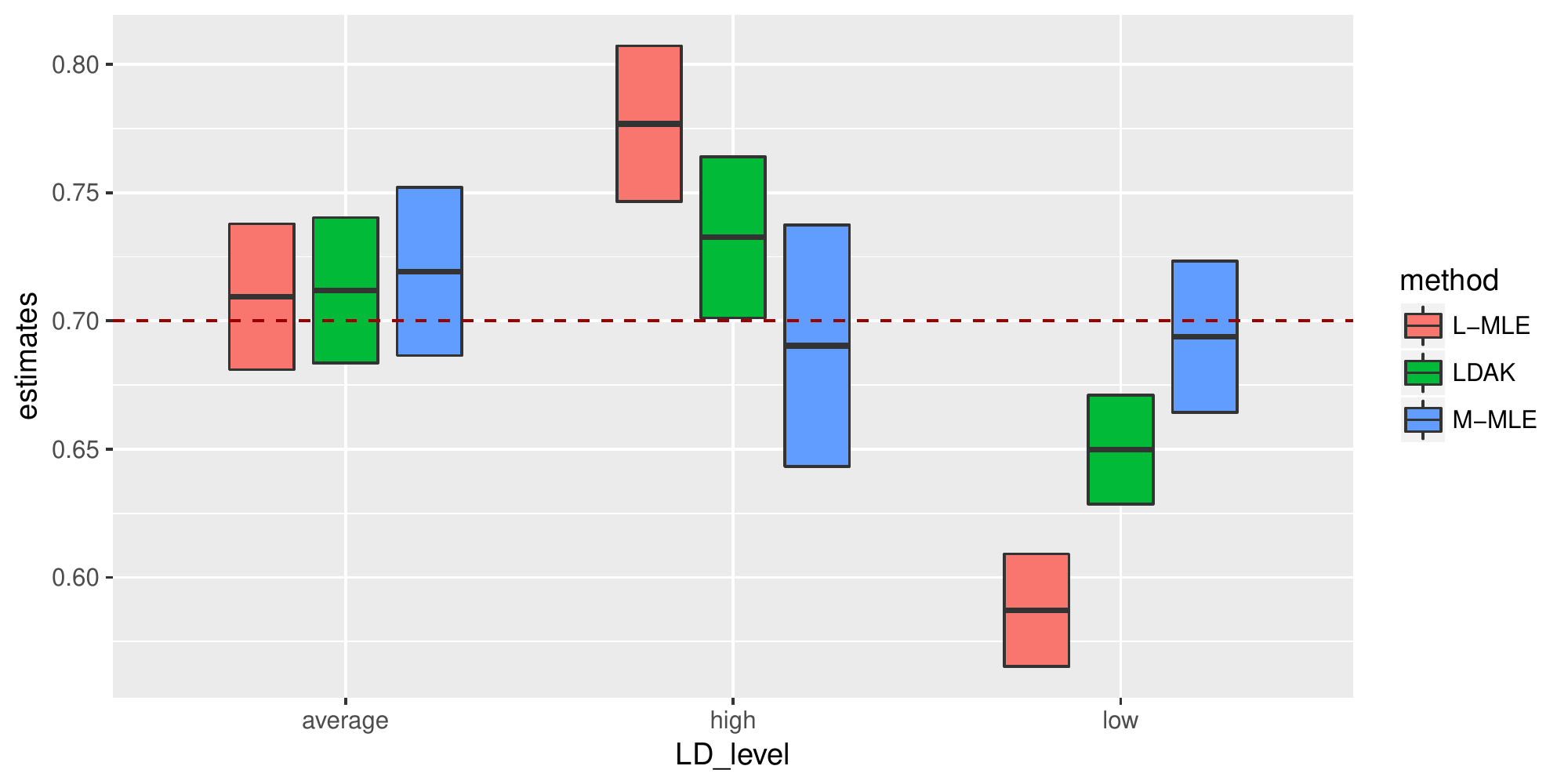}
		\vspace{-0.5cm}
		\caption{95\% confidence intervals of Euclidean (linear) kernel-based maximum likelihood estimator (L-MLE), MLE with LD adjusted Euclideaen GRM (LDAK), and Mahalanobis kernel-based MLE (M-MLE) with causal variants in different different LD regions, based on 50 independent datasets. For LD-level ``average,'' $\mathcal{A}=\mathcal{R}_h\cup \mathcal{R}_l$; ``high,'' $\mathcal{A}=\mathcal{R}_h$; ``low,'' $\mathcal{A}=\mathcal{R}_l$.  Underlying $h^2$ in the top row are $0.3$ and $0.5$ and marked in red dashed line. $h^2$ in the bottom row is 0.7.}\label{fig:h2.full.non-sparse}
		\vspace*{-0.2cm}
	\end{center}
\end{figure}

The set of causal loci $\mathcal{A}$ was chosen in three different ways.  Let $\mathcal{R}_l=\{1,\dots,m/2\}$ and $\mathcal{R}_h=\{m/2+1,\dots,m \}$ be the set of indices indicating low and high LD regions in the simulation setup with within block correlation 0.4 and 0.6, respectively.  In separate simulations, we took the set the causal variants to be 
\begin{itemize}
	\item[(i)]  $\mathcal{A}=\mathcal{R}_h\cup \mathcal{R}_l$;
	\item[(ii)] $\mathcal{A}=\mathcal{R}_h$;
	\item[(iii)] $\mathcal{A}=\mathcal{R}_l$.
\end{itemize}
For each of these settings, we simulated 50 independent datasets and computed heritability estimates.  In particular, for each dataset we computed the Mahalanobis MLE estimator for $h^2$ and the maximum likelihood estimator with Euclidean kernel proposed by \citep{yang2010common,yang2011gcta}. We also used LD-adjusted kinship (LDAK) approach to estimate $h^2$.  LDAK was proposed by \citep{speed2012improved} and is designed to improve the total heritability estimation performance of the Euclidean kernel MLE for $h^2$ when uneven LD structure exists. 
For LDAK, the Euclidean GRM is adjusted by re-weighting each predictor, and the modified REML method takes new inputs $\y$ and LD-adjusted $Z$.  This re-weighting can be viewed as a partial whitening, similar in spirit to the whitening described in Section \ref{sec:proj} for estimating $C$-heritability and the Mahalanobis kernel estimator.  However, the results in this section show that the partial whitening for the LDAK estimator can still lead to biased heritability estimates.  Summary statistics from the simulations are reported in Figure \ref{fig:h2.full.non-sparse}.

In Figure \ref{fig:h2.full.non-sparse}, it's evident that the maximum likelihood estimator with Euclidean kernel is generally biased when causal effects are generated from high or low LD regions. The Euclidean MLE is unbiased when all SNPs are causal (``average'' LD-level in Figure \ref{fig:h2.full.non-sparse}).  In cases where the Euclidean MLE is biased, LDAK has reduced bias, but still some bias remains. The Mahalanobis estimator for $h^2$ is unbiased in all of the settings considered here.  All three of the methods in Figure \ref{fig:h2.full.non-sparse} are maximum likelihood methods; however, the  Mahalanobis MLE has slightly larger standard errors due to the broader distribution of eigenvalues in $\S$, compared to $I$.

\begin{figure}[H]
	\begin{center}
		\vspace{-0.2cm}
		\includegraphics[width=0.49\textwidth]{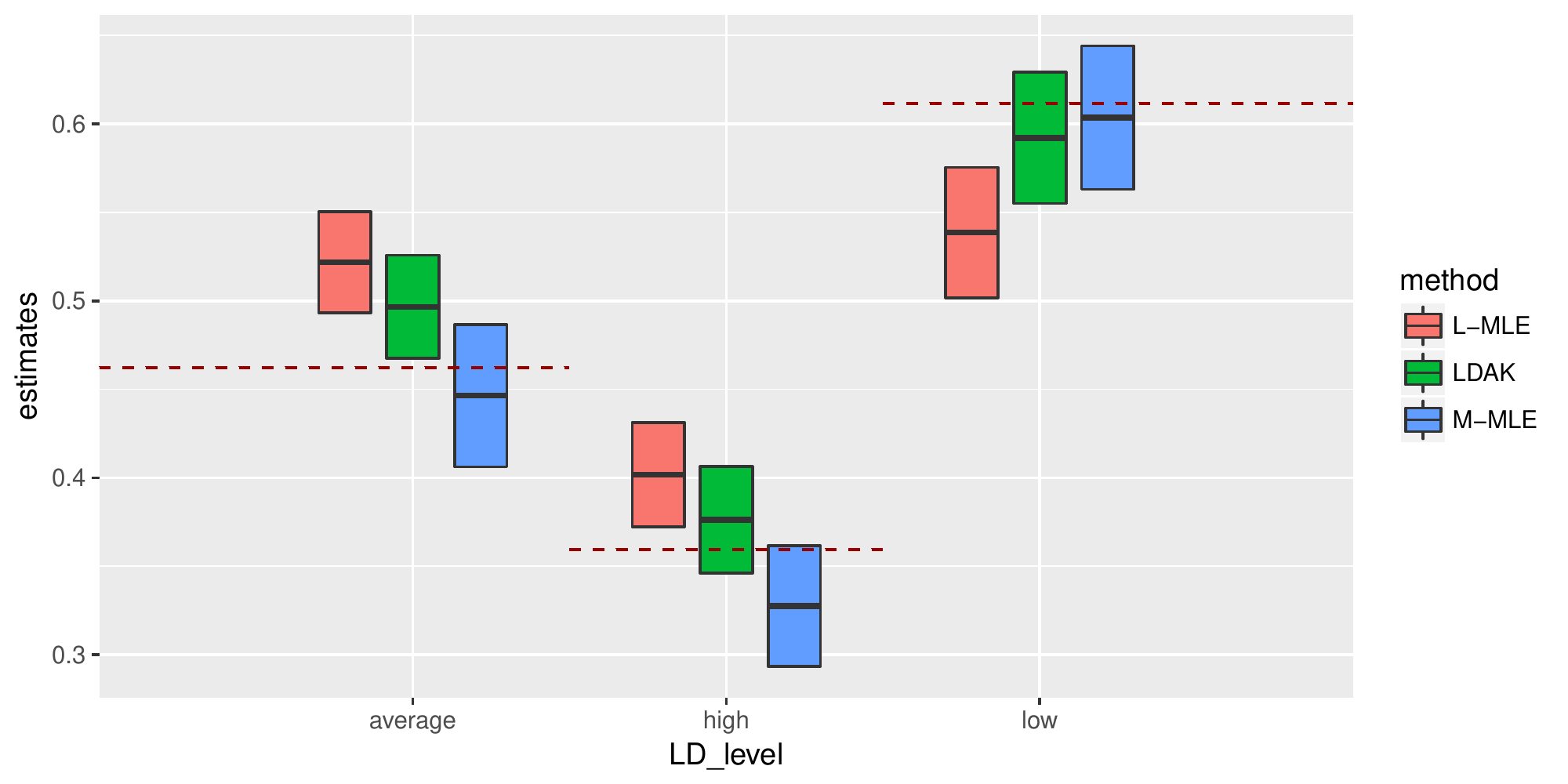}	
		\includegraphics[width=0.49\textwidth]{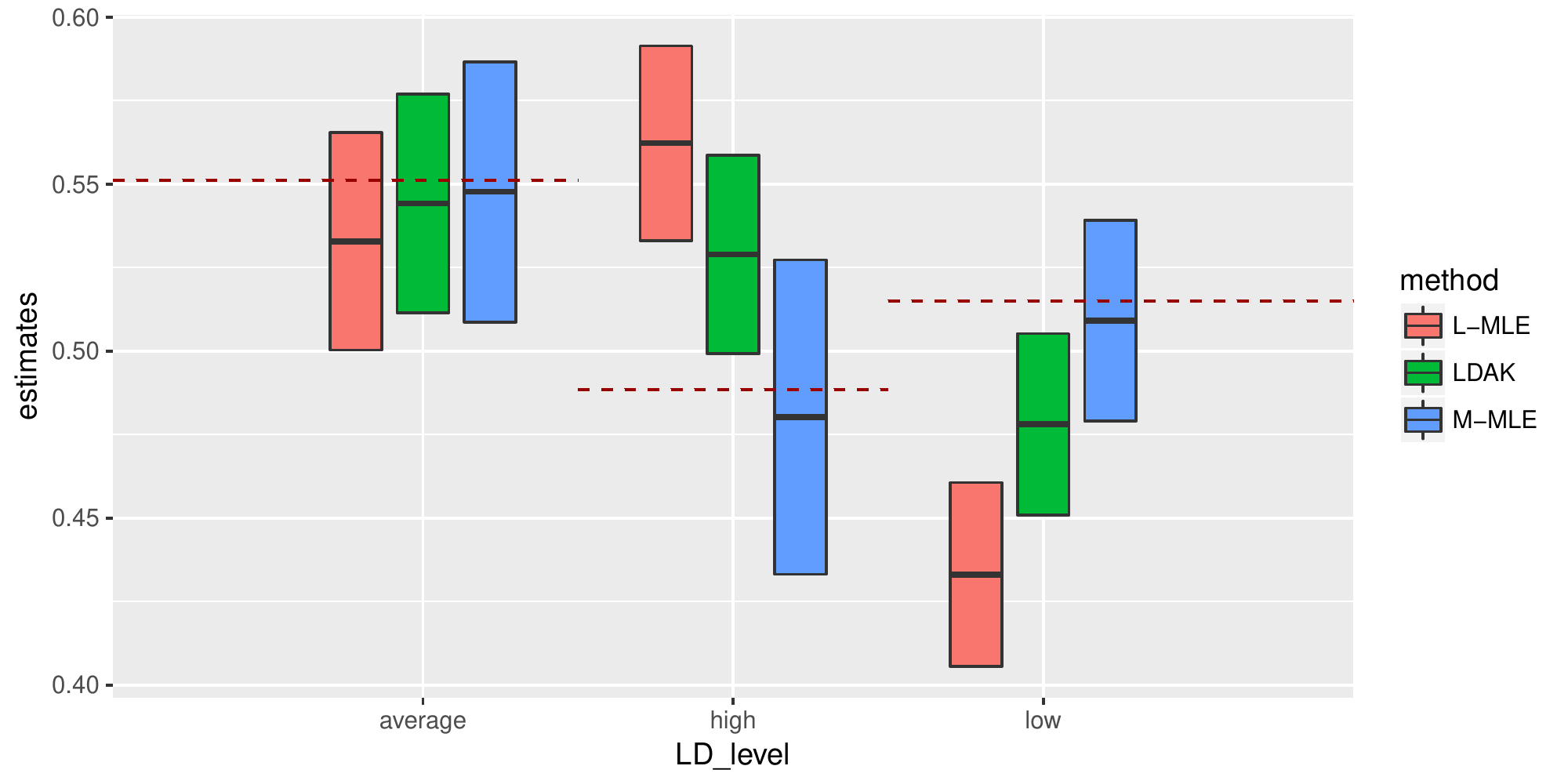}
		\includegraphics[width=0.49\textwidth]{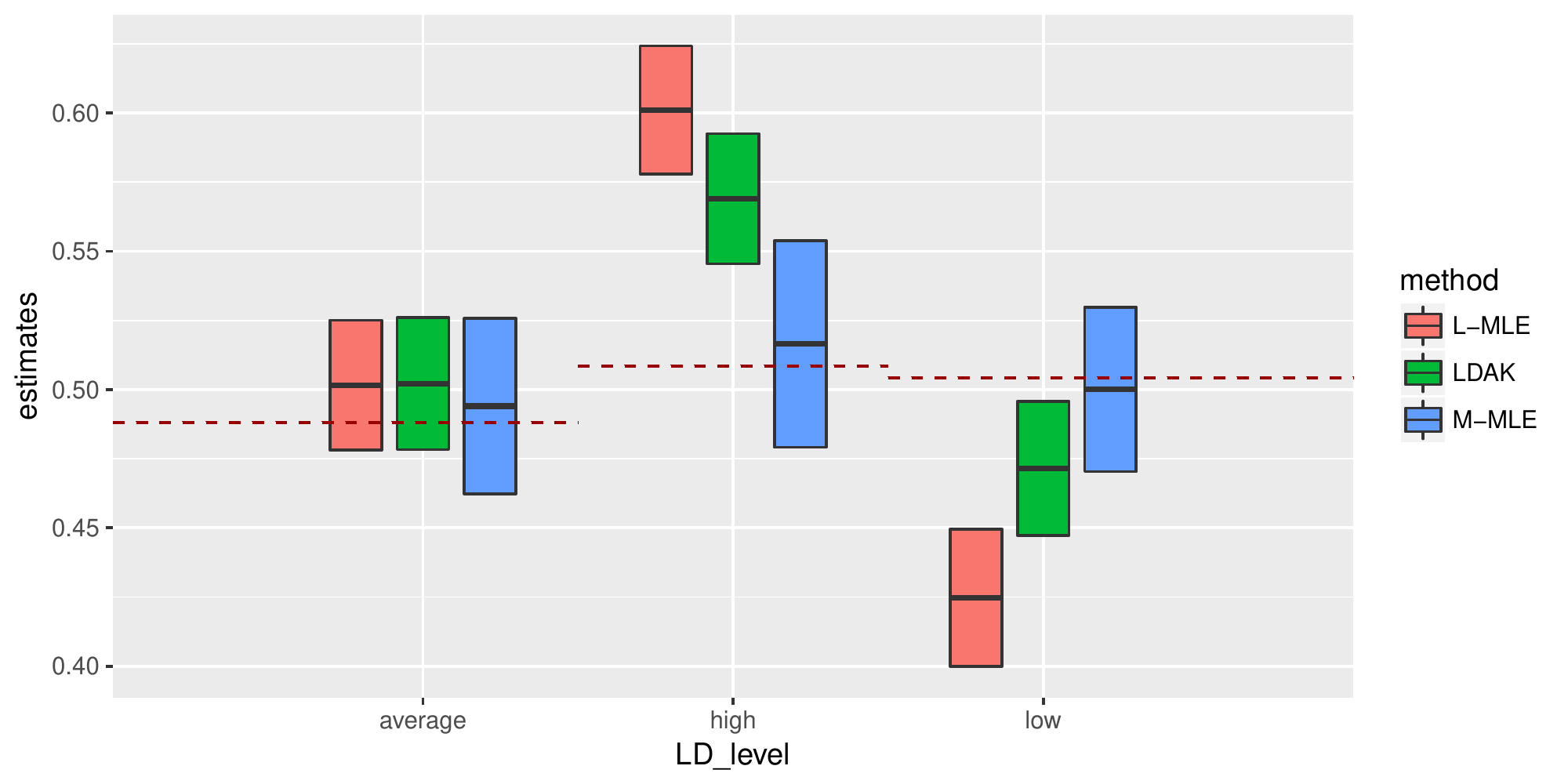}
		\includegraphics[width=0.49\textwidth]{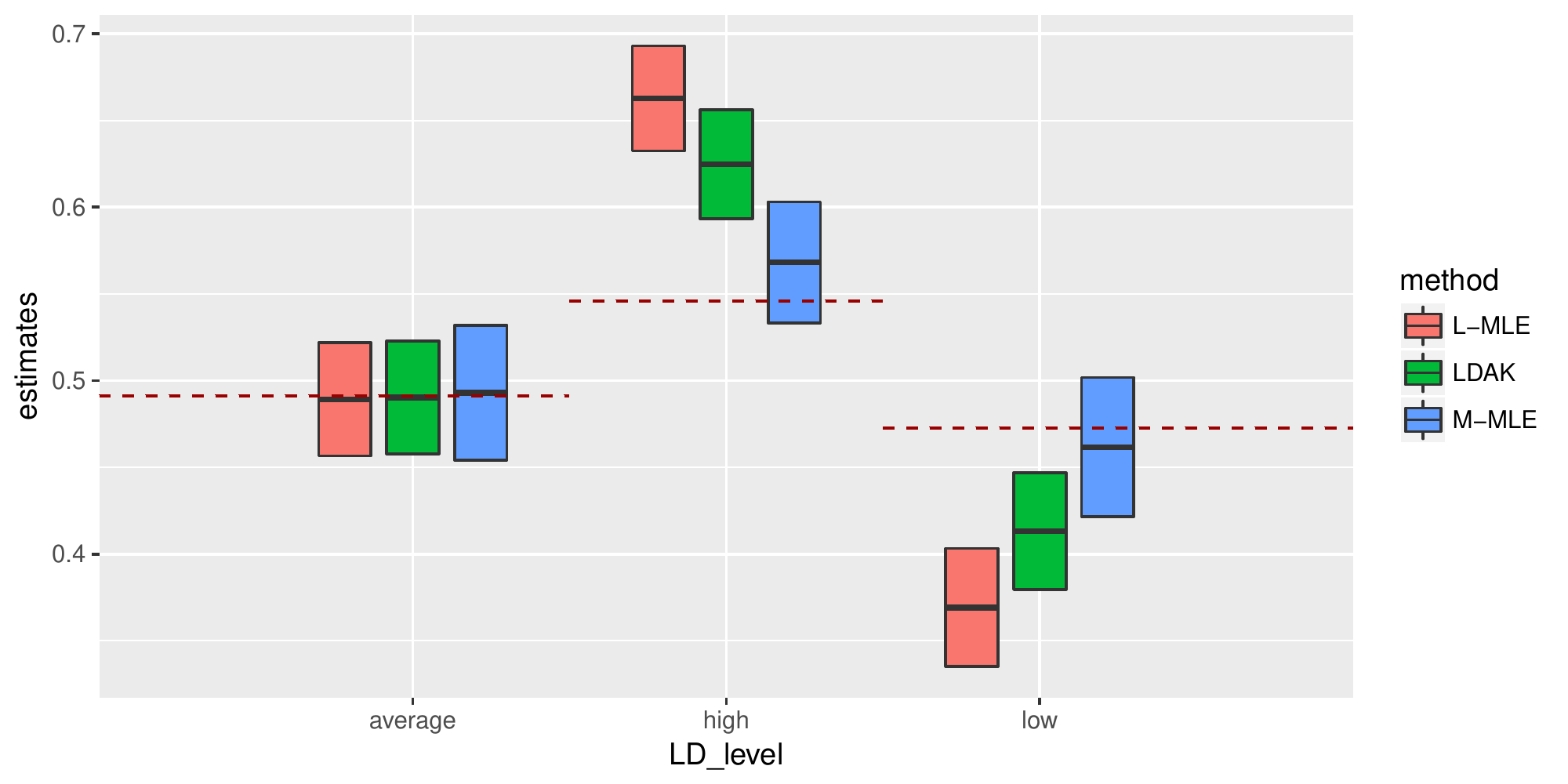}
		\vspace{-0.5cm}
		\caption{95\% Confidence intervals for the Euclidean kernel maximum likelihood estimator (L-MLE), MLE with LD adjusted Euclidean GRM (LDAK), and Mahalanobis kernel-based MLE (M-MLE) with causal variants from different LD-level regions, based on 50 independent datasets. For LD-level ``average,'' $\mathcal{A} \subset\mathcal{R}_h\cup \mathcal{R}_l$; ``high,'' $\mathcal{A} \subset\mathcal{R}_h$; ``low,'' $\mathcal{A}\subset\mathcal{R}_l$.  Underlying $h^2$ is marked in red dashed line. Sparsity (number of causal SNPs) in the top row is $10$ (L) and $50$ (R); sparsity in the bottom row is $200$ (L) and $1,000$ (R). }\label{fig:h2.full.sparse}
		\vspace*{-0.5cm}
	\end{center}
\end{figure}

In a second set of simulations for estimating total heritability, let $\s_g^2=0.5$, we varied the sparsity along with the location of genetic effects. Given $|\mathcal{A}|$, we let $u_j\sim \mathcal N(0,\psi_j)$ for $j\in \mathcal{A}$ with $|\mathcal{A}|=10,50,200$, and $1,000$ and indices in $\mathcal{A}$ sampled uniformly without replacement from the following regions:
\begin{itemize}
	\item[(i)] $\mathcal{A}\subset\mathcal{R}_h\cup\mathcal{R}_l$;
	\item[(ii)]
	$\mathcal{A}\subset\mathcal{R}_h$;
	\item[(iii)]$\mathcal{A}\subset\mathcal{R}_l$.
\end{itemize}
For each of these settigns, we simulated 50 independent datasets with a fixed vector of genetic effects generated from the model (this is the fixed-effects heritability model). For each dataset, we computed three estimates for the heritability coefficient:  The Euclidean MLE, Mahalanobis MLE, and LDAK. Results are shown in Figure \ref{fig:h2.full.sparse}.

\noindent
In Figure \ref{fig:h2.full.sparse}, the underlying $h^2$ varies across LD levels and simulation settings because it depends on $\Sigma$ and the realization of the random causal loci set $\mathcal{A}$ --- recall this is the fixed-effects model. The experiments show that the Euclidean MLE heritability estimator is less stable when effect-sizes become more sparse, which is consistent with results in \citep{speed2012improved}. In Figure \ref{fig:h2.full.sparse}, the Euclidean kernel-based MLE is biased upward when genetic effects are concentrated in high-LD regions, and biased downward when concentrated in low-LD regions.  LDAK mitigates the bias of the Euclidean MLE, but does not completely remove it. The Mahalanobis estimator is unbiased in all settings consider in Figure \ref{fig:h2.full.sparse}.
\subsection{Partitioned heritability estimation}\label{sec:simulations.partitioned}

In this subsection, we consider partitioned heritability and simulate data from a two variance components linear model \eqref{lmm.sim} with
\bes
Z\u = Z_{\mathcal{S}}\u_{\mathcal{S}} + Z_{\mathcal{S}^c} \u_{\mathcal{S}^c}
\ees
and a causal loci model, where $u_j\sim \mathcal{N}(0,\psi_j)$ and 
\bes
\psi_j = \left\{
\begin{array}{ll} 
	\frac{1}{c_\mathcal{S}}\sigma^2_\mathcal{S}(p_j(1-p_j))^{-1}, & \mbox{if } i \in \mathcal{A}_1\subseteq \mathcal{S}, \\
	\frac{1}{c_{\mathcal{S}^c}}\sigma^2_{\mathcal{S}^c}(p_j(1-p_j))^{-1}, & \mbox{if } i \in \mathcal{A}_2 \subseteq\mathcal{S}^c,\\
	0,& \mbox{otherwise}.
	\end{array} \right.
\ees

In the first experiment, we let $\mathcal{A}_1=\mathcal S=\{i\in [m];1\equiv i \ \ (\hbox{mod} 4)\}$ and $\mathcal{A}_2=\mathcal{S}^c$.
We would like to estimate $h^2_\mathcal{S}$, the heritability associated with $\mathcal{S}$, while varying $\sigma^2_{\mathcal{S}}=0.1,0.3,0.5$ and keeping $\sigma^2_e=\sigma^2_{\mathcal{S}}+\s_{\mathcal{S}^c}^2=0.5$. 

\begin{figure}[H]
	\begin{center}
		\includegraphics[width=0.5\textwidth]{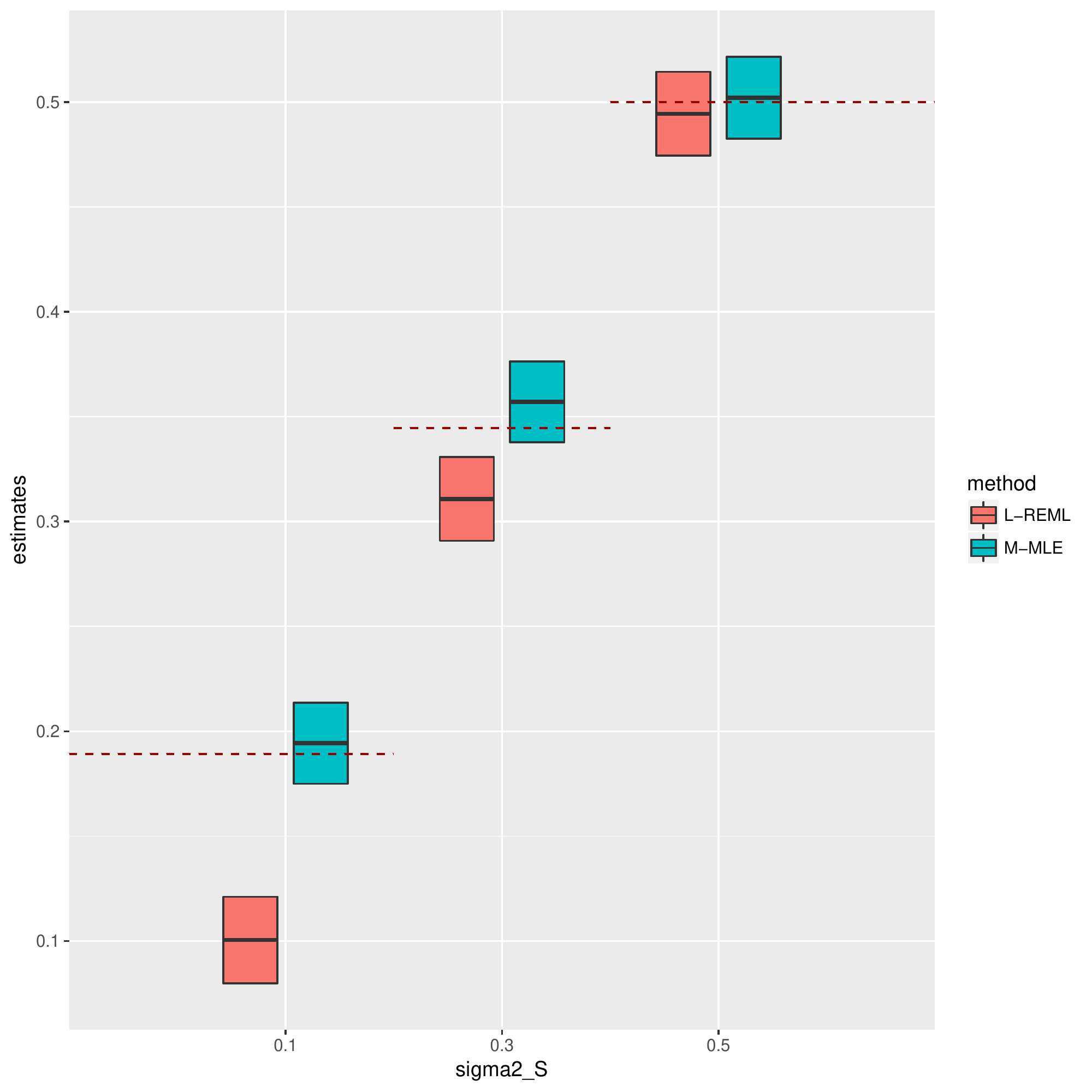}	
		\caption{95\% confidence intervals for partitioned heritability estimation with linear REML (L-REML) and Mahalanobis kernel-based MLE (M-MLE), based on 50 independent datasets.  Signal strength $\sigma^2_\mathcal{S}=0.1,0.3,0.5$ of genetic effects indicated on horizontal axis.  Underlying $h^2_\mathcal{S}$ is marked in red dashed line. }\label{fig:h2.par.LD}
		\vspace*{-0.5cm}
	\end{center}
\end{figure}

For each setting, we simulated 50 independent datasets, and for each dataset we compute the Mahalanobis partitioned heritability estimator and the restricted maximum likelihood (REML) with linear kernel \citep{gilmour1995average,yang2011gcta}. REML finds the maximum likelihood estimator for the two variance components linear model. Summary statistics from these simulations are reported in Figure \ref{fig:h2.par.LD}.

 The Mahalanobis MLE is an unbiased estimator for the partitioned heritability $h^2_\mathcal{S}$ in all settings consider in Figure \ref{fig:h2.par.LD}. The linear REML estimator underestimates $h^2_{\mathcal{S}}$ when $\sigma^2_{\mathcal{S}} < 0.5$.  On the other hand, it's evident that the linear kernel-based MLE is an unbiased estimator for the variance component $\sigma^2_\mathcal{S}$.  In this example, the discrepancy between the REML and Mahalanobis estimators is due to the difference in estimands for the two methods: Under the definition of partitioned heritability in Section \ref{sec:partitioning heritability}, $h^2_{\mathcal{S}} \neq \s_{\mathcal{S}}^2$ in general.  We've argued above that the partitioned heritability coefficient $h^2_{\mathcal{S}}$ correctly accounts for causal loci models and LD. 

\begin{figure}[h]
	\begin{center}
		\includegraphics[width=0.6\textwidth]{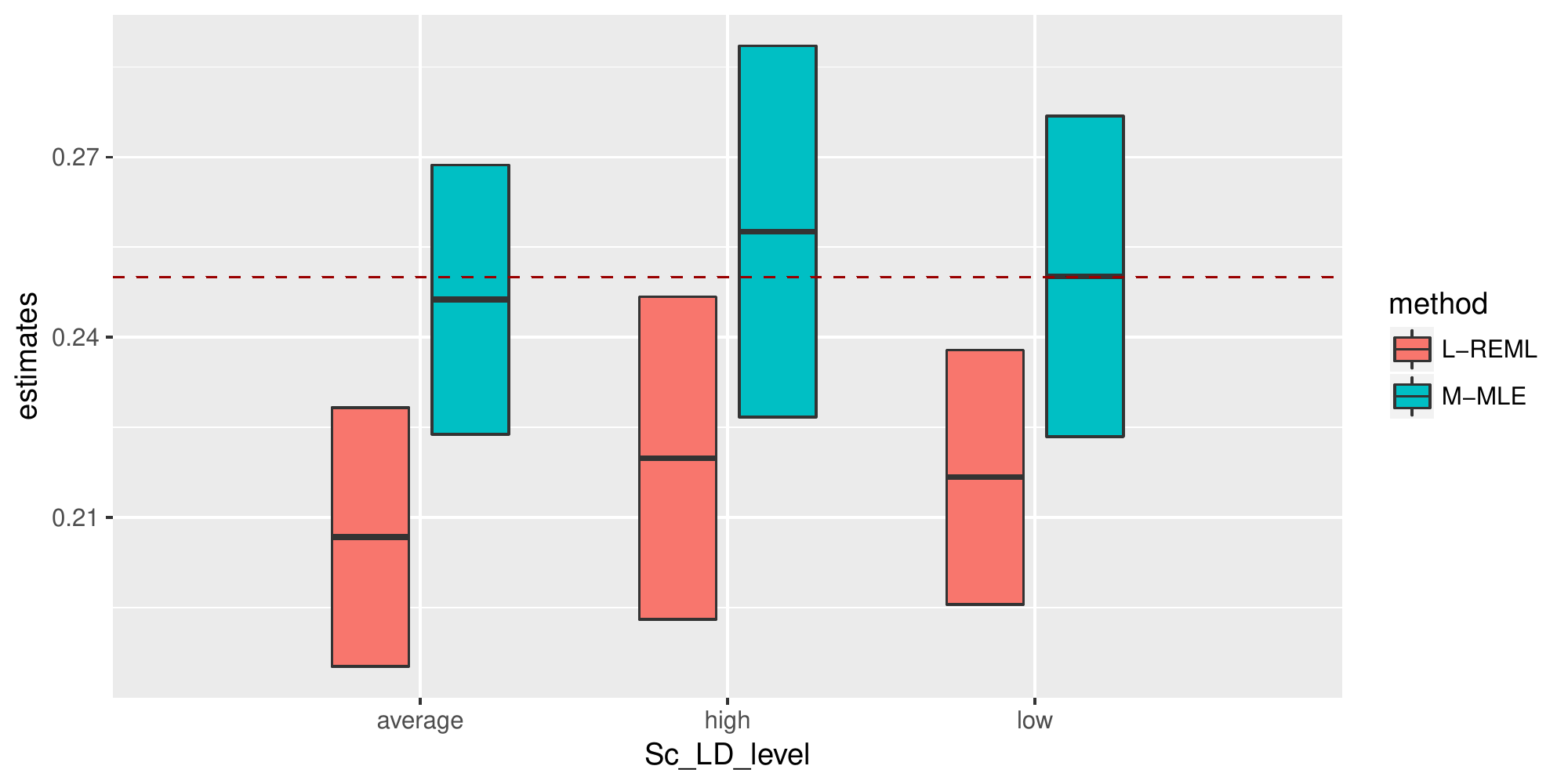}	
		\caption{95\% confidence intervals for partitioned heritability estimation with linear REML (L-REML) and Mahalanobis kernel-based MLE (M-MLE), based on 50 independent datasets. Underlying $h^2_\mathcal{S}$ is marked in red dashed line. Causal loci $\mathcal{S}$ are locat in low-LD region, while causal loci in $\mathcal{S}^c$ varies are from regions with varying LD.}\label{fig:h2.par.causal}
		\vspace*{-0.5cm}
	\end{center}
\end{figure}

In the next experiment, we set  
\begin{itemize}
	\item[(i)] $\mathcal{S}=\{m/4+1,\dots,3m/4\}$,
	\item [(ii)] $\mathcal{A}_1=\{m/4+1,m/2\}$,
	\item [(iii)] $\sigma^2_\mathcal{S}=0.25$ and $\sigma^2_e=0.5$.
	
\end{itemize}
Let $\mathcal{R}_l=\{1,\dots,m/4\}$ and $\mathcal{R}_h=\{3m/4+1,\dots,m \}$ be the set of indices corresponding to low and high LD regions in $\mathcal{S}^c$, respectively. 
In this experiment we varied the location of $\mathcal{A}_2$, so that 
\begin{itemize}
	\item [(i)] $\mathcal{A}_2=\mathcal{R}_l\cup\mathcal{R}_h$;
	\item[(ii)]  $\mathcal{A}_2=\mathcal{R}_h$;
	
	\item [(iii)]  $\mathcal{A}_2=\mathcal{R}_l$.
	
\end{itemize}
For each simulation setting, we generated 50 random-effects vectors and independent datasets; for each dataset we computed the Mahalanobis estimator and the REML estimator with linear kernel for the partitioned heritability due to $\mathcal{S}$.
Summary statistics are reported in Figure \ref{fig:h2.par.causal}.

In all three settings depicted in Figure \ref{fig:h2.par.causal}, the Mahalanobis estimator is an unbiased estimator for 
\bes
h^2_\mathcal{S}=\frac{\sigma^2_\mathcal{S}}{\sigma^2_\mathcal{S}+\sigma^2_{\mathcal{S}^c}+\sigma^2_e}=0.25,
\ees 
while the linear REML estimator has downward bias. Note that under this causal loci model, the linear REML estimators are also biased for estimating the variance components; see Table \ref{par.sim1}.  This is due to model misspecifiction -- REML assumes that all genetic effects in $\mathcal{S}$ and $\mathcal{S}^c$ are iid, but in this causal loci model half of the genetic effects in $\mathcal{S}$ equal 0, i.e. $u_j = 0$ for $j \in \mathcal{S}\setminus\mathcal{A}_1$.  

\begin{table}[H]
	\begin{center}
		\caption{Estimated mean (95\% confidence interval) of $\sigma^2_\mathcal{S}$, $\sigma^2_{\mathcal{S}^c}$ and $\sigma^2_{e}$ by linear REML. Based on results from 50 independent datasets.}
		\label{par.sim1}
		\begin{tabular}{c|lc|lc|lc}
			 & \multicolumn{2}{|c|}{Average LD Level} &  \multicolumn{2}{|c|}{High LD Level} & \multicolumn{2}{|c}{Low LD Level} \\ \hline
			\multirow{2}{0.6in}{$\sigma^2_\mathcal{S}=0.25$} & Mean:&  0.206   & Mean:& 0.219  &  Mean:& 0.218  \\
			&95\%CI:& (0.184,0.228)&95\%CI:&(0.192,0.247)&95\%CI:&(0.194,0.240) \\ \hline
			\multirow{2}{0.6in}{$\sigma^2_{\mathcal{S}^c}=0.25$}  &Mean:& 0.235&Mean:& 0.292 &Mean:&	0.199\\
			&95\%CI:&(0.210,0.260)&95\%CI:&(0.274,0.309)&95\%CI:& (0.175,0.223)\\ \hline
			\multirow{2}{0.6in}{$\sigma^2_{e}=0.5$}  & Mean:&0.554 & Mean:&0.486\ &Mean:&0.579\\
			&95\%CI:&(0.526,0.582)&95\%CI:&(0.457,0.516)&95\%CI:&(0.554,0.604)
		\end{tabular}
	\end{center}
\end{table}

\section{Discussion}\label{sec:discussion}
In this paper, we studied the Mahalanobis distanced-based GRM for estimating heritability with GWAS data. Under the Mahalanobis kernel, the fixed- and random-effects model are equivalent, which  resolves many LD-related inconsistencies in total and partitioned heritability estimation at the modeling level. This paper also re-emphasizes the importance of understanding the underlying LD structure as the LD matrix is required for computing the Mahalanobis distance.  An interesting research direction is to study semi-supervised learning methods for estimating the LD structure from unlabeled genotype data (with no corresponding phenotypes) and using this to improve heritability estimation for a specific phenotype of interest; this has connections with recent theoretical work in statistics on the conditionality principle in high dimensions \citep{azriel2018conditionality}.

Theoretically, the connection between fixed- and random-effects models with the Mahalanobis kernel is most clearly delineated assuming multivariate Gaussian genotypes. However, the simulation results suggest that the proposed estimator performs reliably for genetic data with non-Gaussian genotypes. In this paper, only quantitative traits are considered. An interesting research direction is extending the Mahalanobis MLE results to binary traits in heritability estimation through either the liability or generalized linear model. 

\appendix

\section*{Appendix}

\noindent {\bf Proof of Proposition \ref{prop:ph}.}

Without loss of generality, assume that $\Var(y)=1$ and $\mathcal{S}=\{1,\dots,|\mathcal{S}|\}$.  Let $\u=(\u_\mathcal{S}^\T,\u_{\mathcal{S}^c}^\T)^\T \in \mathcal{R}^p$ and $\S=\begin{pmatrix}
	\S_{\mathcal{S}}& \S_{\mathcal{S},\mathcal{S}^c}\\
	\S_{\mathcal{S},\mathcal{S}^c}^\T& \S_{\mathcal{S}^c}.
	\end{pmatrix}$. Then the quadratic form based heritability $h^2_\mathcal{S}(\u,\S)=\u^\T \Gamma\u$ where $
	\G=\begin{pmatrix}
	\G_{\mathcal{S}}& \G_{\mathcal{S},\mathcal{S}^c}\\
	\G_{\mathcal{S},\mathcal{S}^c}^\T& \G_{\mathcal{S}^c}
	\end{pmatrix}$ is a $p\times p$ matrix. Moreover, due to property (i), $0\leq\G\leq \S$. If $\u_\mathcal{S}^c=0$, $\forall\ \u_\mathcal{S}\in \mathcal{R}^{|\mathcal{S}|}$,
	\bes
	h^2&=&\u_\mathcal{S}^\T \S_{\mathcal{S}}\u_\mathcal{S},
	\cr h^2_\mathcal{S}&=&\u_\mathcal{S}^\T\G_{\mathcal{S}}\u_\mathcal{S}.
	\ees
	By property (ii), this implies $\G_{\mathcal{S}}=\S_\mathcal{S}$. 
	If $\u_\mathcal{S}=0$ and $\u_{\mathcal{S}^c}\neq 0$, 
	\bes
	h^2&=&\u_{\mathcal{S}^c}^\T\S_{\mathcal{S}^c}\u_{\mathcal{S}^c},
	\cr h^2_\mathcal{S}&=&\u_{\mathcal{S}^c}^\T\G_{\mathcal{S}^c}\u_{\mathcal{S}^c}.
	\ees
	Then by Property (i) and (ii), $\G_{\mathcal{S}^c}<\S_{\mathcal{S}^c}$. Next, Property (i) suggests that 
	\bes
	\S-\G=
	\begin{pmatrix}
		0& \S_{\mathcal{S},\mathcal{S}^c}-\G_{\mathcal{S},\mathcal{S}^c}\\
		(\S_{\mathcal{S},\mathcal{S}^c}-\G_{\mathcal{S},\mathcal{S}^c})^\T& \S_{\mathcal{S}^c}-\G_{\mathcal{S}^c}
	\end{pmatrix}\geq 0
	\ees
	Since $\S_{\mathcal{S}^c}-\G_{\mathcal{S}^c}>0$, $\S-\G\geq 0$ is equivalent to
	\bes
	0-(\S_{\mathcal{S},\mathcal{S}^c}-\G_{\mathcal{S},\mathcal{S}^c})(\S_{\mathcal{S}^c}-\G_{\mathcal{S}^c})^{-1}(\S_{\mathcal{S},\mathcal{S}^c}-\G_{\mathcal{S},\mathcal{S}^c})^\T&\geq& 0
	\cr(\S_{\mathcal{S},\mathcal{S}^c}-\G_{\mathcal{S},\mathcal{S}^c})(\S_{\mathcal{S}^c}-\G_{\mathcal{S}^c})^{-1}(\S_{\mathcal{S},\mathcal{S}^c}-\G_{\mathcal{S},\mathcal{S}^c})^\T&\leq& 0
	\ees
	However, $(\S_{\mathcal{S},\mathcal{S}^c}-\G_{\mathcal{S},\mathcal{S}^c})(\S_{\mathcal{S}^c}-\G_{\mathcal{S}^c})^{-1}(\S_{\mathcal{S},\mathcal{S}^c}-\G_{\mathcal{S},\mathcal{S}^c})^\T\geq 0$ because $\S-\G\geq 0$
	and $\S_{\mathcal{S}^c}-\G_{\mathcal{S}^c}>0$. Therefore, 
	\bes
	(\S_{\mathcal{S},\mathcal{S}^c}-\G_{\mathcal{S},\mathcal{S}^c})(\S_{\mathcal{S}^c}-\G_{\mathcal{S}^c})^{-1}(\S_{\mathcal{S},\mathcal{S}^c}-\G_{\mathcal{S},\mathcal{S}^c})^\T&=&0
	\ees 
	Thus, $\G_{\mathcal{S},\mathcal{S}^c}=\S_{\mathcal{S},\mathcal{S}^c}$. Moreover, $\G\geq 0$ implies
	\bel{geq}
	\G_{\mathcal{S}^c}-\S_{\mathcal{S},\mathcal{S}^c}^\T\S_{\mathcal{S}}^{-1} \S_{\mathcal{S},\mathcal{S}^c}\geq 0
	\eel
	We then let 
	\bes
	\G_{\mathcal{S}^c}=\S_{\mathcal{S},\mathcal{S}^c}^\T\S_{\mathcal{S}}^{-1} \S_{\mathcal{S},\mathcal{S}^c}+M, \quad  M \geq 0
	\ees
	Finally, we would like to prove $M=0$ by contradiction. Suppose that there exist some $\u_{\mathcal{S}^c}=\bb$ and $\S$ such that $\bb^\T M\tilde \bb>0$. Let $\u =(0, \dots, 0, \ \bb)^\T$, then 
	\bes
	h^2_\mathcal{S}=\bb^\T\S_{\mathcal{S},\mathcal{S}^c}^\T\S_{\mathcal{S}}^{-1} \S_{\mathcal{S},\mathcal{S}^c}\bb+\bb^\T M \bb 
	\ees
	Now let 
	\bes
	\tilde \S=\begin{pmatrix}
		\S_\mathcal{S}& \S_{\mathcal{S},\mathcal{S}^c}\\
		\S_{\mathcal{S},\mathcal{S}^c}^\T& \S_{\mathcal{S},\mathcal{S}^c}^\T\S_\mathcal{S}^{-1}\S_{\mathcal{S},\mathcal{S}^c}+\dfrac{1}{2}M+\dfrac{\bb^\T M\bb}{4\|\bb\|_2^2}I
	\end{pmatrix}>0
	\ees
	By property (iii), $h^2_\mathcal S(\u,\tilde\S)=h^2_\mathcal S(\u, \S)$.
	However, 
	\bes
	h^2(\u,\tilde{\S})&=&\bb^\T\S_{\mathcal{S},\mathcal{S}^c}^\T\S_{\mathcal{S}}^{-1} \S_{\mathcal{S},\mathcal{S}^c}\bb+\frac{3}{4}\bb^\T M \bb
	\cr &=&h^2_\mathcal{S}(\u,\tilde \S)-\frac{1}{4}{\bb}^\T M\bb
	\ees
	This contradicts with Property (i). Therefore, $M=0$ and $\G_{\mathcal{S}^c}=\S_{\mathcal{S},\mathcal{S}^c}^\T\S_{\mathcal{S}}^{-1} \S_{\mathcal{S},\mathcal{S}^c}$.

\hfill $\Box$

\noindent {\bf Proof of Proposition \ref{ch-mle}.}

The consistency part of Proposition \ref{ch-mle} follows immediately from Theorem 1 of \citep{dicker2016maximum}.  By Theorem 2 of \citep{dicker2016maximum},
	\bes
	\sqrt{n}(\hat\eta^2_C-\eta^2_C)\overset{\mathcal{D}}{\longrightarrow}\mathcal{N}(0,\psi).
	\ees
	where $\psi=(\iota_2-\iota_3^2/\iota_4)^{-1}$ and 
	\bes
	\iota_\alpha=\frac{1}{2n\sigma_{C^\perp}^{2(4-\alpha)}}\tr\left\{\left(\frac{1}{k}W_CW_C^\T\right)^{\alpha-2}\left(\frac{\eta^2_C}{k}W_CW_C^\T+I\right)^{2-\alpha}
	\right\}.
	\ees
	Let $\mathcal I=\frac{1}{k}W_CW_C^\T$ and $\mathcal J=\frac{\eta^2_C}{k}W_CW_C^\T+I$,
	It follows that 
	\bes
	\psi=2\sigma^4_{C^\perp}\left(1-\frac{\tr(\mathcal I\mathcal J^{-1})^2}{n\tr(\mathcal{I}^2\mathcal{J}^{-2})}\right)^{-1}.
	\ees
	By the Delta method,
	\bes
	\sqrt{n}(\hat{h}^2_C-h^2)\overset{\mathcal D}{\longrightarrow}\mathcal{N}\left(0,\frac{\psi}{(1+\eta^2_C)^4}\right).
	\ees 
\hfill $\Box$
\bibliographystyle{imsart-nameyear}
\bibliography{reference.bib}

\end{document}